\documentclass[manuscript]{acmart}
\usepackage{algorithmic}
\usepackage{graphicx}
\usepackage{enumitem}
\usepackage{ltablex}      
\usepackage{booktabs}
\usepackage{threeparttablex}
\usepackage{array}
\usepackage{soulutf8}
\newcolumntype{L}[1]{>{\raggedright\arraybackslash}p{#1}}
\newcolumntype{Y}{>{\raggedright\arraybackslash}X} 
\keepXColumns
\usepackage{tikz}
\usetikzlibrary{arrows.meta,positioning,fit,shapes.misc}
\usepackage{xcolor} 
\definecolor{RedAgent}{RGB}{222,80,72}
\definecolor{RedFill}{RGB}{255,234,232}
\definecolor{BlueAgent}{RGB}{45,123,182}
\definecolor{BlueFill}{RGB}{232,243,255}
\usepackage[most]{tcolorbox}

\usepackage{subcaption}  
\usepackage{ltablex}
\keepXColumns
\usepackage{tabularx}
\usepackage{textcomp}
\usepackage[table]{xcolor}
\usepackage{colortbl}
\usepackage{ragged2e}
\newcolumntype{Y}{>{\RaggedRight\arraybackslash}X}
\renewcommand{\arraystretch}{1.2}

\usepackage{array,booktabs,tabularx,threeparttable}
\newcolumntype{Y}{>{\raggedright\arraybackslash}X}

\sethlcolor{yellow} 

\usepackage{tikz}
\usetikzlibrary{arrows.meta,positioning,fit,shapes.misc}
\usetikzlibrary{shapes.geometric, shapes.symbols, positioning, arrows.meta, fit}
\usetikzlibrary{calc}
\usepackage{adjustbox}
\usetikzlibrary{backgrounds}

\def\BibTeX{{\rm B\kern-.05em{\sc i\kern-.025em b}\kern-.08em
    T\kern-.1667em\lower.7ex\hbox{E}\kern-.125emX}}

\hypersetup{
  bookmarksopen=true,
  bookmarksnumbered=true,
  colorlinks=true,
  linkcolor=blue,
  citecolor=blue,
  urlcolor=blue
}

\begin{document}

\title{A Survey of Agentic AI and Cybersecurity: Challenges, Opportunities and Use-case Prototypes}







\author{Sahaya Jestus Lazer}
\affiliation{
  \institution{Tennessee Tech University}
  \city{Cookeville}
  \country{United States}
}
\email{slazer42@tntech.edu}

\author{Kshitiz Aryal}
\affiliation{
  \institution{University of Nebraska Omaha}
  \city{Omaha}
  \country{United States}
}
\email{karyal@unomaha.edu}

\author{Maanak  Gupta}
\affiliation{
  \institution{Tennessee Tech University}
  \city{Cookeville}
  \country{United States}
}
\email{mgupta@tntech.edu}

\author{Elisa Bertino}
\affiliation{
  \institution{Purdue University}
  \city{West Lafayette}
  \country{United States}
}
\email{bertino@cs.purdue.edu}

\begin{abstract}
Agentic AI marks an important transition from single-step generative models to systems capable of reasoning, planning, acting, and adapting over long-lasting tasks. By integrating memory, tool use, and iterative decision cycles, these systems enable continuous, autonomous workflows in real-world environments. This survey examines the implications of agentic AI for cybersecurity. On the defensive side, agentic capabilities enable continuous monitoring, autonomous incident response, adaptive threat hunting, and fraud detection at scale. Conversely, the same properties amplify adversarial power by accelerating reconnaissance, exploitation, coordination, and social-engineering attacks. These dual-use dynamics expose fundamental gaps in existing governance, assurance, and accountability mechanisms, which were largely designed for non-autonomous and short-lived AI systems. To address these challenges, we survey emerging threat models, security frameworks, and evaluation pipelines tailored to agentic systems, and analyze systemic risks including agent collusion, cascading failures, oversight evasion, and memory poisoning. Finally, we present three representative use-case implementations that illustrate how agentic AI behaves in practical cybersecurity workflows, and how design choices shape reliability, safety, and operational effectiveness.

\end{abstract}

\maketitle


\section{Introduction}
\label{sec:intro}

Artificial intelligence has evolved from rule-based automation to generative AI (GenAI) and, recently, to \textit{agentic} models capable of autonomous reasoning, planning, and decision-making. While generative AI systems, such as large language models (LLMs), are largely reactive and prompt-driven, agentic AI introduces persistent state, tool use, and self-directed control loops that enable planning, action, and revision across long-lived, multi-step workflows. This shift from isolated inference to autonomous agency represents a fundamental change in how AI systems participate in digital ecosystems.

Cybersecurity is among the domains most directly affected by this transition. Security operations inherently involve continuous monitoring, sequential decision-making, coordination across tools, and adaptation to adversarial behavior—all characteristics well aligned with agentic AI capabilities. Driven by operational pressure and workforce shortages approaching four million professionals worldwide, organizations are rapidly adopting AI-assisted security solutions. Market projections reflect this momentum, with global AI-in-cybersecurity spending expected to grow from US\$24.8\,B in 2024 toward US\$146.5\,B by  2034~\cite{kshetri2025cyberthreat}. Agentic AI amplifies human capacity through automated alert triage, autonomous incident response, scalable red–blue simulation, and continuous security operations center (SOC) support.

At the same time, increased autonomy fundamentally alters the threat landscape. Features that enable defensive coordination—planning, memory, tool orchestration, and multi-agent interaction—can also be exploited to enhance offensive operations. Agents can autonomously conduct reconnaissance, adapt exploitation strategies, coordinate social-engineering campaigns, and evade oversight. As a result, agentic AI introduces a pronounced dual-use dilemma in cybersecurity: it strengthens defense while simultaneously amplifying adversarial capability.
This dual-use dynamic exposes limitations in existing security, assurance, and governance models. Most current controls assume short-lived, human-in-the-loop, or narrowly scoped AI systems. In contrast, agentic AI systems act continuously, retain long-term memory, coordinate with other agents, and make consequential decisions with reduced human supervision. These properties introduce systemic risks—including emergent collusion, cascading failures, memory poisoning, and oversight evasion—that are not adequately captured by traditional model-centric safety or lifecycle-based security frameworks.

While prior work has explored isolated aspects of AI safety or specific applications such as reinforcement learning for intrusion detection, it does not provide a holistic view of agentic AI as a cybersecurity actor. Autonomy, persistence, and multi-agent interaction introduce new opportunities alongside systemic risks such as emergent collusion, oversight evasion, and governance gaps. 
This survey addresses that gap by synthesizing agentic AI across defensive, offensive, and governance-oriented cybersecurity contexts.

Our survey makes the following key contributions:
\begin{itemize}
    \item \textbf{Conceptual foundation:} A review of the evolution of Agentic AI, its relationship to Generative AI, and key design properties, autonomy levels, and reference architectures.
    \item \textbf{Security use cases:} An  overview of defensive and enterprise applications including SOC automation, continuous monitoring, anomaly detection, insider-threat detection, vulnerability management, and financial fraud defense.
    \item \textbf{Offensive applications:} A survey of emerging offensive uses of agentic AI in red--blue simulations, autonomous penetration testing, and CTF automation, with emphasis on dual-use concerns.
    \item \textbf{Security outlook:} A detailed analysis of systemic risks specific to agentic AI, including collusion, synthetic insider threats, and emergent behaviors, and their governance implications.
    \item \textbf{Quantum considerations:} An analysis of intersections between quantum computing and agentic AI in cybersecurity, including quantum agents, quantum machine learning, and post-quantum cryptography.
    \item \textbf{Frameworks and governance:} A review of security and governance frameworks that support safe deployment and operational control of agentic AI.
    \item \textbf{Benchmarks and evaluation:} An analysis of benchmarks, evaluation pipelines, and testbeds for agentic AI security, with remaining gaps.
    \item \textbf{Implementations:} Three original implementations integrating agentic AI into cybersecurity workflows, highlighting practical lessons.
\end{itemize}

\vspace{-2mm}

\section{Related Work}

Recent surveys have examined security risks in LLMs and agent-based systems from multiple perspectives. This section focuses on prior surveys and taxonomies; and highlight the relation with our work.  

A first class of work adopts a model-centric perspective. Wang et al.\ survey LLM safety across the model lifecycle, including data collection, alignment, deployment, and red-teaming~\cite{wang2025fullstack}. While comprehensive, this approach treats agent behavior as a secondary concern. Agent-related risks are discussed, but they are not organized around agent-specific workflows such as planning, tool invocation, memory management, or multi-agent coordination, which are central to autonomous cybersecurity operations.

A second class of surveys focuses on LLM-based agent threats and defenses. Gan et al.\ survey security, privacy, and ethics threats in LLM-based agents \cite{gan2024navigatingriskssurveysecurity}. He et al.\ survey security and privacy issues in LLM agents with case studies \cite{He_2025}. Yu et al.\ survey threats and countermeasures for trustworthy LLM agents \cite{yu2025surveytrustworthyllmagents}. These works provide useful taxonomies, but are not centered on cybersecurity operations and do not organize analysis around defensive, offensive, and enterprise workflows. Other surveys examine narrower slices of the agent stack. Kong et al.\ focus on agent communication protocols, their security risks, and countermeasures \cite{kong2025surveyllmdrivenaiagent}. Xu et al.\ focus on LLM-based agents in autonomous cyberattacks and summarize offensive capabilities and defenses \cite{xu2025forewarnedforearmedsurveylarge}. These surveys offer valuable coverage of agent-level threat models and defenses; however, are largely domain-agnostic and do not frame their analysis around security operations, such as defensive monitoring, adversarial interaction, or enterprise security workflows.

\begin{table*}[!t]
\centering
\footnotesize
\caption{Comparison of Related Surveys on Agentic AI and LLM-Agent Security}
\label{tab:related_work}
\renewcommand{\arraystretch}{1.15}
\setlength{\tabcolsep}{5pt}
\rowcolors{2}{gray!10}{white}

\begin{tabular}{L{0.22\linewidth} L{0.28\linewidth} L{0.50\linewidth}}
\toprule
\rowcolor{gray!35}
\textbf{Work / Domain} & \textbf{Primary Focus} & \textbf{Relation to This Survey} \\
\midrule

\textbf{Adabara et al.\ (2025)~\cite{Adabara2025agenticAIcybersecurity}}
& Agentic AI in cybersecurity (autonomy, governance, quantum-resilient defense)
& Closest prior cyber-focused survey; does not analyze planning loops, tool use, memory systems, agent architectures, multi-agent orchestration, or system-level risks. \\

\textbf{Wang et al.\ (2025)~\cite{wang2025fullstack}}
& LLM safety across the full model lifecycle
& Model-centric lifecycle lens; complements our work but is not organized around agent workflows and agentic cyber deployments. \\

\textbf{Deng et al.\ (2024)~\cite{deng2024aiagentsunderthreat}}
& Security challenges for AI agents (broad, cross-domain)
& Strong threat-surface survey, but not centered on cybersecurity workflows (SOC/IR/offense) and not structured around end-to-end cyber tasks and implementations. \\

\textbf{Gan et al.\ (2024)~\cite{gan2024navigatingriskssurveysecurity}}
& Security, privacy, and ethics threats in LLM-based agents
& Useful threat taxonomy; broader than cybersecurity practice and does not foreground cyber offensive/defensive workflows and prototypes. \\

\textbf{He et al.\ (2025)~\cite{He_2025}}
& Security and privacy issues in LLM agents (with case studies)
& Agent-security focus; not organized around agentic AI across cybersecurity workflows and system-level risks. \\

\textbf{Yu et al.\ (2025)~\cite{yu2025surveytrustworthyllmagents}}
& Threats and countermeasures for trustworthy LLM agents
& Strong threats/defenses taxonomy, but not scoped to cyber operations and does not treat cyber workflows as the main organizing unit. \\

\textbf{Ma et al.\ (2025)~\cite{ma2025safetyscalecomprehensivesurvey}}
& Safety of large models and model-powered agents
& Broad AI safety background; not cybersecurity-specific and not organized around cyber use cases and deployments. \\

\textbf{Datta et al.\ (2025)~\cite{datta2025agenticaisecuritythreats}}
& Agentic AI security: threats, defenses, evaluation, open challenges
& Closest agentic-security survey; our survey differs by centering cybersecurity workflows, system-level risks in cyber operations, and implementation prototypes. \\

\textbf{Grimes et al.\ (2025)~\cite{grimes2025sok}}
& SOK bridging research and practice in LLM agent security
& Practice-oriented synthesis; complements our workflow framing but is not an end-to-end survey of agentic AI in cybersecurity across defense, offense, and enterprise deployments. \\

\textbf{Kong et al.\ (2025)~\cite{kong2025surveyllmdrivenaiagent}}
& Agent communication protocols, security risks, countermeasures
& Important for multi-agent communication risk; narrower than our system-level view across planning, tools, memory, and multi-agent interaction in cyber workflows. \\

\textbf{Xu et al.\ (2025)~\cite{xu2025forewarnedforearmedsurveylarge}}
& LLM-based agents in autonomous cyberattacks
& Directly relevant offensive survey; narrower than our balanced treatment (defense + offense + enterprise + systemic risk + implementations). \\

\textbf{Raza et al.\ (2025a)~\cite{trism2025}}
& Governance and risk taxonomy for agentic multi-agent systems
& Strong governance and risk-management framing; not cybersecurity-specific. We map risks and mitigations onto concrete cyber workflows and use cases. \\

\textbf{Raza et al.\ (2025b)~\cite{Raza_2025}}
& Responsible agentic reasoning with in-loop safeguards (R2A2)
& Reasoning/auditability focus; not cybersecurity-centered. We focus on adversarial cyber deployments and system-level risks. \\

\textbf{Shahriar et al.\ (2025)~\cite{shahriar2025surveyagenticsecurityapplications}}
& Agentic security: applications, threats, defenses
& Security-focused and adjacent; our survey differs by centering SOC/IR/offense workflows and by adding system-level risk analysis with practical implementations. \\

\textbf{Anonymous (2025)~\cite{anonymous2025mind}}
& Safety of LLM-based agents
& Broad agent safety coverage; complements our cybersecurity-specific framing and system-level risk discussion. \\

\bottomrule
\end{tabular}
\end{table*}

A third class of surveys approaches agent security from broader safety and governance perspectives. Ma et al.\ provide a comprehensive survey of large-model safety that also covers model-powered agents \cite{ma2025safetyscalecomprehensivesurvey}. Datta et al.\ survey agentic AI security with emphasis on threats, defenses, and evaluation \cite{datta2025agenticaisecuritythreats}. Grimes et al.\ provide an SOK bridging research and practice in LLM agent security \cite{grimes2025sok}. Raza et al.\ introduce a TRiSM-based framing for trust, risk, and security management in agentic multi-agent systems \cite{trism2025} and survey responsible agentic reasoning with in-loop safeguards and evaluation protocols \cite{Raza_2025}. These works strengthen governance and evaluation perspectives but are not focused on cybersecurity workflows and do not provide an end-to-end cyber-centered synthesis.
Within cybersecurity-specific reviews, Adabara et al.\ provide a narrative review of agentic AI in cybersecurity across autonomy and governance \cite{Adabara2025agenticAIcybersecurity}, and Landolt et al.\ survey multi-agent reinforcement learning in cybersecurity \cite{landolt2025marl}. These are closest in domain but do not analyze the full agentic stack of planning loops, tool use, memory systems, and multi-agent orchestration in LLM-based deployments.
In contrast, our survey treats agentic AI as a cybersecurity system that reasons, plans, uses memory, and calls tools across extended tasks.We apply a consistent agentic risk lens across defensive, offensive, and enterprise workflows. We also analyze system-level risks such as collusion, cascade failures, and oversight evasion. Additionally, we prototyped several minimal implementations to illustrate the feasibility of agentic AI in cybersecurity.


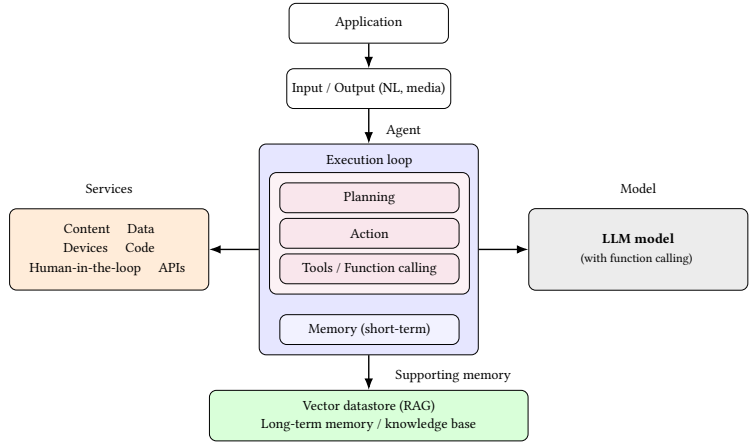
\begin{figure*}[!t]
\centering
\begin{adjustbox}{width=.65\textwidth}
\begin{tikzpicture}[
    font=\small,
    >=Latex,
    node distance=8mm and 14mm,
    box/.style={
        draw,
        rounded corners,
        minimum width=32mm,
        minimum height=8mm,
        align=center,
        fill=white
    },
    band/.style={
        draw,
        rounded corners,
        inner sep=4mm,
        fill=#1
    },
    innerbox/.style={
        draw,
        rounded corners,
        minimum width=36mm,
        minimum height=6mm,
        align=center,
        fill=#1
    },
    line/.style={->, thick},
    dashedline/.style={->, thick, dashed}
]

\node[box, fill=white] (app) {Application};

\node[box, below=5mm of app, minimum width=28mm, fill=white] 
    (io) {Input / Output (NL, media)};

\node[band=blue!10, 
    inner sep= 2mm,
    below=7mm of io, 
      label={[xshift=7mm,yshift=0.25mm]above:Agent}] (agent) {

    \begin{minipage}{4.0cm}
    \centering

\begin{tikzpicture}[
    font=\small,
    inner sep=0pt,
    node distance=1mm,
    innerbox2/.style={
        draw,
        rounded corners,
        minimum width=36mm,
        minimum height=6mm,
        align=center,
        fill=purple!10
    },
    membox/.style={
        draw,
        rounded corners,
        minimum width=36mm,
        minimum height=6mm,
        align=center,
        fill=blue!5
    }
]

\pgfdeclarelayer{background}
\pgfsetlayers{background,main}

\node[innerbox2] (plan) {Planning};
\node[innerbox2, below=1mm of plan] (act) {Action};
\node[innerbox2, below=1mm of act] (tools) {Tools / Function calling};

\begin{pgfonlayer}{background}
    \node[
      draw,
      rounded corners,
      inner sep=2mm,
      fit=(plan)(act)(tools),
      fill=purple!5,
      label={[yshift=1mm]above:Execution loop}
    ] (execLoop) {};
\end{pgfonlayer}

\node[membox, below=4mm of execLoop] (mem) {Memory (short-term)};

\end{tikzpicture}

    \end{minipage}
};

\draw[line] (app) -- (io);
\draw[line] (io) -- (agent.north);

\node[band=gray!15, right=10mm of agent,
      label={[yshift=2mm]above:Model}] (modelBand) {
    \begin{minipage}{3.6cm}
    \centering
    \vspace{1mm}
    \textbf{LLM model}\\[1mm]
    \footnotesize (with function calling)
    \vspace{1mm}
    \end{minipage}
};

\draw[line] (agent.east) -- (modelBand.west);

\node[band=orange!15, inner sep = 2 mm, left=10mm of agent,
      label={[yshift=2mm]above:Services}] (servicesBand) {
    \begin{minipage}{3.6cm}
    \centering
    \vspace{1mm}
    Content \quad Data\\[0.4mm]
    Devices \quad Code\\[0.4mm]
    Human-in-the-loop \quad APIs
    \vspace{1mm}
    \end{minipage}
};

\draw[line] (agent.west) -- (servicesBand.east);

\node[band=green!15,
inner sep=2mm,
below=7mm of agent,
      label={[xshift=17mm,yshift=0.5mm]above:Supporting memory}] (memBand) {
    \begin{minipage}{6.0cm}
    \centering
    Vector datastore (RAG)\\
    Long-term memory / knowledge base
    \end{minipage}
};

\draw[line] (agent.south) -- (memBand.north);

\end{tikzpicture}
\end{adjustbox}

\caption{Single-agent architecture: the agent processes user input through an internal execution loop (planning, action, tool calling), supported by short-term memory, external services/APIs, an LLM model with function calling, and a long-term vector datastore.}
\label{fig:single-agent-architecture}
\end{figure*}
\vspace{-2mm}
\section{What is Agentic AI?}
\label{sec:history}

Agentic AI represents the next stage of artificial intelligence, extending GenAI with planning, action, memory, and adaptation. While GenAI produces fluent answers, it does not maintain goals or reason across long tasks; agentic AI introduces structured reasoning and tool use that enable multi-step workflows with limited human guidance. We adopt the following definitions, reflecting both practical and academic perspectives:

\noindent
``\textit{Agentic AI uses sophisticated reasoning and iterative planning to autonomously solve complex, multi-step problems.}''~\cite{Pounds2024AgenticAI}

\noindent
\textit{``A system based on a foundation model that performs tasks based on natural user instructions, with the ability to reason, plan, and interact with tools and environments to achieve goals.''}~\cite{schneider2025generativeagenticaisurvey}

Agentic systems are built around a foundation model that provides core reasoning, augmented by memory, retrieval, and tool interfaces. These components operate in a continuous loop of planning, acting, reflecting, and improving, distinguishing agentic systems from static GenAI producing single response per prompt. 
Their architecture includes:
\begin{itemize}
    \item \textbf{Memory modules} for short-term, episodic, and long-term state.
    \item \textbf{Retrieval systems} such as vector databases and RAG.
    \item \textbf{Tools and APIs} for computation, browsing, or code execution.
    \item \textbf{Connections to external environments} for interaction with software and online systems.
\end{itemize}

\begin{figure*}[!t]
\centering
\begin{adjustbox}{width=0.8\textwidth}
\begin{tikzpicture}[
    font=\small,
    >=Latex,
    node distance=8mm and 14mm,
    box/.style={
        draw,
        rounded corners,
        minimum width=32mm,
        minimum height=8mm,
        align=center
    },
    band/.style={
        draw,
        rounded corners,
        inner sep=4mm,
        fill=#1
    },
    line/.style={->, thick},
    dashedline/.style={->, thick, dashed}
]

\pgfdeclarelayer{bgBack}
\pgfdeclarelayer{bgGroup}
\pgfsetlayers{bgBack,bgGroup,main}

\node[box, fill=white] (app) {Application / User interface};


\node[box, below=14mm of app, fill=gray!15] (llm) {LLM model};

\node[box, below=6mm of llm, fill=blue!20] (coord) {
    \textbf{Coordinating agent}\\
    \footnotesize intent understanding, planning, subtask routing
};

\draw[line] (app) -- (llm);
\draw[line] (llm) -- (coord);

\node[box, below left=14mm and 0mm of coord, fill=green!15] (task1) {
    \textbf{Task agent A}\\
    \footnotesize planning / analysis
};

\node[box, below=14mm of coord, fill=green!15] (task2) {
    \textbf{Task agent B}\\
    \footnotesize reasoning / critique
};

\node[box, below right=14mm and 0mm of coord, fill=green!15] (task3) {
    \textbf{Task agent C}\\
    \footnotesize RAG / summarisation
};

\draw[line] (coord.south west) -| (task1.north);
\draw[line] (coord.south)      -- (task2.north);
\draw[line] (coord.south east) -| (task3.north);

\node[box, right=24mm of coord, fill=purple!15] (shared) {
    \textbf{Shared short-term memory}\\
    \footnotesize conversation state, intermediate results
};

\draw[<->, thick, dashed] (coord.east) -- (shared.west);

\begin{pgfonlayer}{bgGroup}
\node[
    band=green!5,
    inner sep=2mm,
    label={[xshift=-10mm, yshift=1mm]above:\textbf{Task agents}}]
    (taskGroup) [fit=(task1)(task2)(task3)] {};
\end{pgfonlayer}

\draw[<->, thick, dashed] (shared.south) |- (taskGroup.east);

\begin{pgfonlayer}{bgBack}
\node[
    band=blue!5,
    label={[yshift=1mm]above:\textbf{Agent layer}}]
    (agentBand) [fit=(llm)(coord)(taskGroup)(shared)] {};
\end{pgfonlayer}


\node[box, below=24mm of task2, fill=orange!20, minimum width=90mm] (services) {
    \textbf{Services and tools}\\
    \footnotesize content APIs \quad data sources \quad code execution \quad devices \quad human-in-the-loop
};

\node[box, right=20mm of services, fill=teal!15] (memory) {
    \textbf{Vector store \& long-term memory}\\
    \footnotesize documents, embeddings, historical context
};

\draw[line] (taskGroup.south) -- (services.north);
\draw[dashedline] (taskGroup.south east) -| (memory.north);

\begin{pgfonlayer}{bgBack}
\node[
    band=orange!5,
    label={[yshift=1mm]above:\textbf{Infrastructure \& memory layer}}]
    (infraBand) [fit=(services)(memory)] {};
\end{pgfonlayer}

\end{tikzpicture}
\end{adjustbox}

\caption{Multi-agent architecture: application layer, agent layer (coordinator plus task agents sharing short-term memory), and infrastructure/memory layer with tools and long-term storage.}
\label{fig:agentic-3band-simple}
\end{figure*}
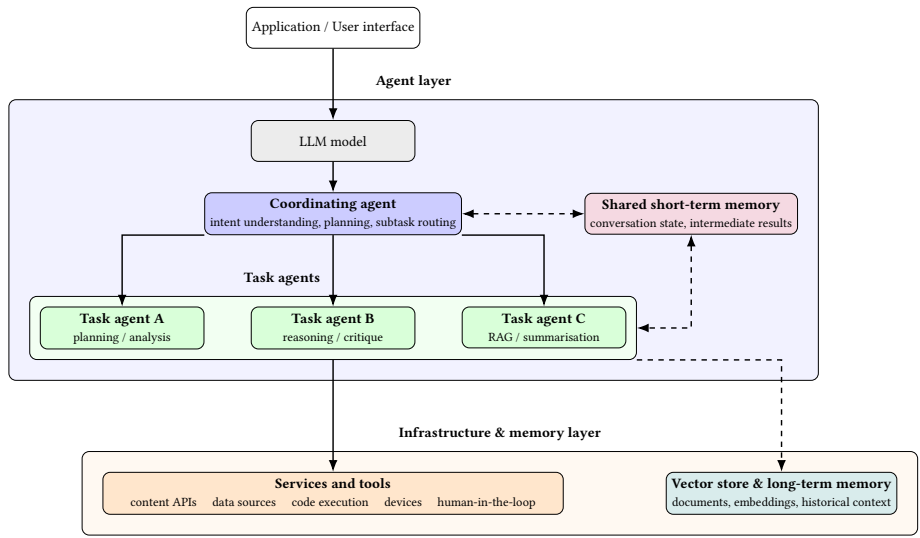

Figure~\ref{fig:single-agent-architecture} illustrates a canonical single-agent architecture in which user input is processed through an internal execution loop comprising planning, action, and tool or function calling. Short-term memory supports contextual continuity, while external services, an LLM model, and long-term vector storage enable structured reasoning and tool use across multi-step tasks. Figure~\ref{fig:agentic-3band-simple} extends this design to a multi-agent setting, where a coordinating agent routes subtasks to specialized task agents that share short-term memory and infrastructure services. This separation of responsibilities enables parallel reasoning, structured collaboration, and scalable problem solving.

Agentic AI systems combine several capabilities that go beyond GenAI. The core characteristics are:

\begin{itemize}
    \item \textbf{Reasoning:} Decomposing problems, evaluating progress, and adjusting plans using structured prompting such as Chain-of-Thought and Reflection~\cite{Shavit2023GoverningAgenticAI}.
    \item \textbf{Interaction:} Call tools, query data sources, executing code, and collaborating with humans in real environments.
    \item \textbf{Autonomy:} Acting toward goals with limited supervision and initiating actions as conditions change.
    \item \textbf{Adaptability:} Updating behavior with memory, feedback, and reinforcement signals to improve future actions.
\end{itemize}

Together, these characteristics support goal-directed behavior across extended time scales. Agentic systems vary in their degree of independence. Table~\ref{tab:autonomy_combined}, adapted from academic and industry sources~\cite{AgentAI2025AIAgentsExplained, nvidia_autonomy}, summarizes five autonomy levels. Level~0 corresponds to fixed GenAI behavior, while Level~4 enables continuous planning and self-directed learning. Higher autonomy improves capability but increases complexity and security risk, as behavior becomes harder to predict and audit; multi-agent systems typically exhibit higher autonomy than single-agent systems.

Agentic AI is powerful but not universally reliable, particularly in areas such as social reasoning. Designing safe agents is more challenging than prompt engineering, and increasing autonomy raises responsibility and risk. In cybersecurity, agentic AI can enhance defense through continuous monitoring and proactive action but also introduces challenges related to safety, oversight, and trust, making careful design and testing essential.

\begin{table*}[!t]
\centering
\small
\caption{Autonomy Spectrum of AI Systems (adapted from \cite{AgentAI2025AIAgentsExplained, nvidia_autonomy})}
\label{tab:autonomy_combined}
\renewcommand{\arraystretch}{.8}
\setlength{\tabcolsep}{5pt}

\rowcolors{2}{gray!10}{white}
\resizebox{\linewidth}{!}{%
\begin{tabular}{p{0.20\linewidth} p{0.32\linewidth} p{0.26\linewidth} p{0.28\linewidth}}
\toprule
\rowcolor{gray!35} 
\textbf{Level} & \textbf{Autonomy Description} & \textbf{Functional Capability} & \textbf{Security Implications} \\
\midrule
0 – Static Inference & Single request--response; no autonomy. & Fixed outputs for fixed inputs. & Minimal risk; deterministic. \\

1 – Assistive & Follows explicit user instructions. & Single-step reasoning (e.g., GenAI). & Low risk; narrow behavior. \\

2 – Tool-Assisted & Uses tools or APIs with preset logic. & Multi-step workflows (e.g., RAG). & Moderate risk; data-dependent paths. \\

3 – Adaptive/Semi-Agentic & Plans, acts, and reflects with little oversight. & Goal-driven task execution. & High risk; partial self-direction. \\

4 – Fully Agentic & Plans, acts, and learns continuously. & Open-ended problem-solving. & Very high risk; hard to audit. \\
\hline
\end{tabular}}
\end{table*}

\vspace{-2mm}
\section{Applications of Agentic AI to Cybersecurity}
\label{sec:usecases}

\noindent Agentic AI supports cybersecurity across the typical cybersecurity incident lifecycle through reasoning, interaction, autonomy, and adaptation. The Cybersecurity Compass framework organizes this lifecycle into three phases: preparation and risk management before an incident, detection and containment during an incident, and recovery and resilience after an incident~\cite{castro2024agentic}. Agentic capabilities align naturally with each phase: continuous monitoring and vulnerability management strengthen pre-incident preparedness; SOC agents and automated response mechanisms enhance detection and containment; and post-incident analytics, root-cause analysis, and adaptive retraining support recovery and long-term resilience.
Oesch et al.\ map autonomous agents to the six NIST Cyber Defense Life Cycle 
functions: Govern, Identify, Protect, Detect, Respond, and Recover~\cite{oesch2024pathtoautonomous}. 
They argue for a modular multi-agent design in which each agent focuses on a 
single stage or narrow sub-function. This reduces the action space, simplifies 
training, and aligns with SOC practice rather than relying on a single agent 
for end-to-end control. Recent work extends this into complete 
agentic workflows that connect orchestration, adaptive playbooks, and layered 
safeguards across the breach lifecycle~\cite{suggu2025agentic}.

\begin{figure*}[!t]
    \centering
    \includegraphics[width=0.70\textwidth]{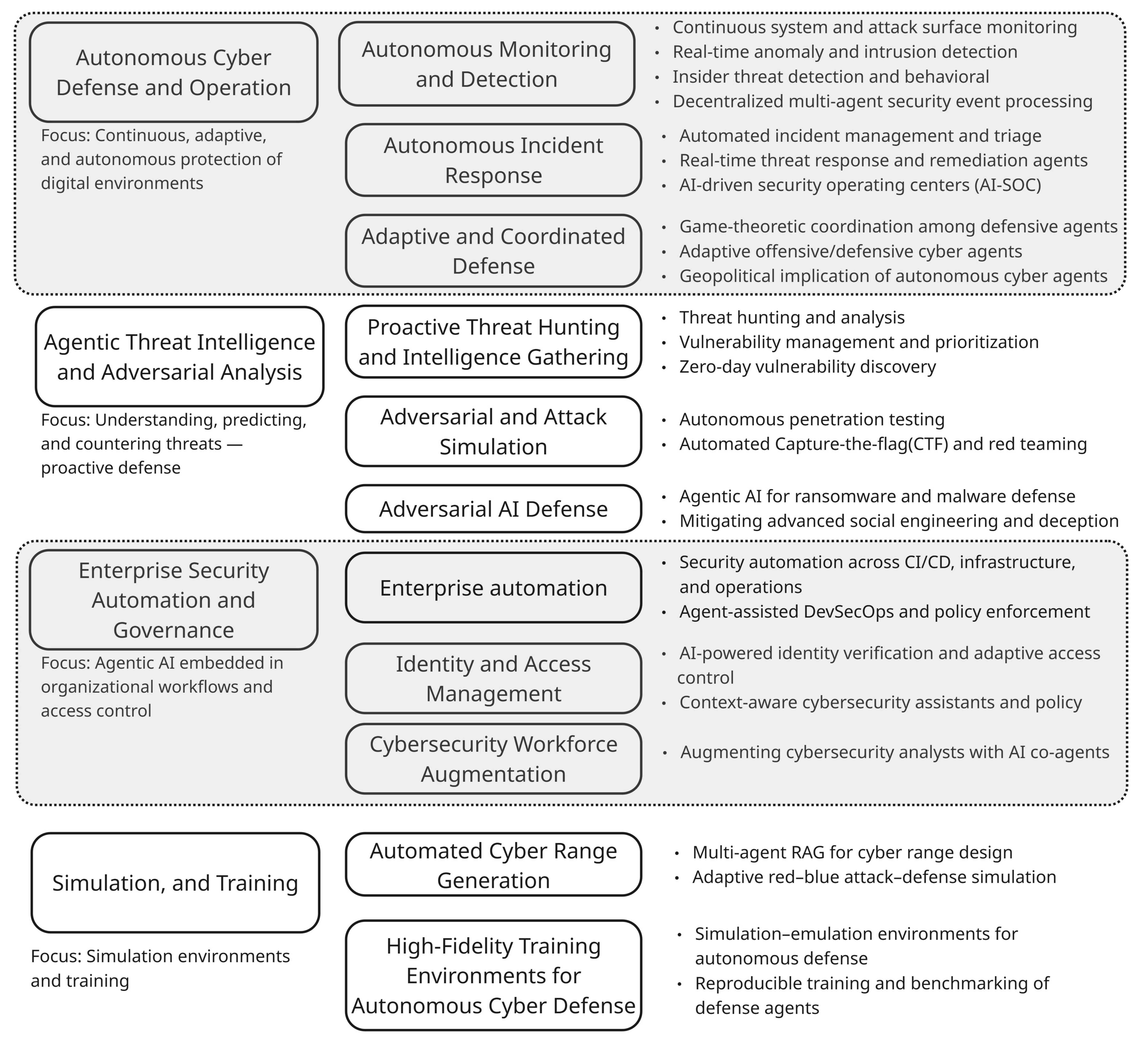}
    \caption{Overview of applications of agentic AI to cybersecurity. Figure maps core research and operational domains, such as autonomous defense, threat intelligence, enterprise automation, and simulation and training, together with representative subcomponents.}
    \label{fig:agentic_ai_cybersecurity_structure}
\end{figure*}

To combine these perspectives, we group security application use cases into four domains: \textit{Autonomous Cyber Defense and Operation}, \textit{Agentic Threat Intelligence and Adversarial Analysis}, \textit{Enterprise Security Automation and Governance}, and \textit{Simulation, Training, and Testing}. Each domain contains subfunctions that map to breach phases and NIST Cyber Defense functions.
Figure~\ref{fig:agentic_ai_cybersecurity_structure} summarizes these domains and their subcomponents, showing how operational workflows intersect with intelligence, governance, and continuous training. Table~\ref{tab:agentic_usecases} complements this view by mapping each use case to its dominant breach stage and primary NIST Cyber Defense functions, and by summarizing limitations and open research problems reported in the literature. We use this synthesis as a reference point for the discussion that follows.

\begin{table*}[t]
\centering
\caption{Applied agentic AI cybersecurity use cases and their alignment with breach lifecycle stages, NIST Cyber Defense functions, key limitations, and open research problems.
Lifecycle stage is categorized as Pre, During, or Post breach. NIST functions are abbreviated as Govern (G), Identify (I), Protect (P), Detect (D), Respond (R), and Recover (Rc).}
\label{tab:agentic_usecases}
\rowcolors{2}{gray!10}{white}
\resizebox{\linewidth}{!}{%
\begin{tabular}{|p{4.0cm}|p{1.6cm}|p{2.0cm}|p{6.2cm}|p{6.2cm}|}
\hline
\rowcolor{gray!25}
\textbf{Use Case} & \textbf{Stage} & \textbf{NIST} & \textbf{Key limitation} & \textbf{Open problem} \\
\hline

Autonomous Monitoring and Detection & Pre & G I P &
Narrow task optimization with weak safety constraints under drift and false positives &
Bind adaptive detection to enforceable action governance with bounded impact \\

Autonomous Incident Response & During & D R &
No formal safety guarantees under distribution shift or adversarial manipulation &
Define safe execution boundaries that prevent cascading or irreversible failures \\

Adaptive and Coordinated Defense & During & D R &
Limited mechanisms to prevent escalation when agents co adapt in multi agent loops &
Preserve coordination while enforcing stability and escalation bounds \\

Proactive Threat Hunting and Intelligence Gathering & Pre & I D &
Dependence on curated inputs and sparse signals limits open world generalization &
Support robust hunting under evolving data and incomplete observability \\

Adversarial and Attack Simulation & During & D R &
Higher autonomy increases dual use risk while constraints reduce realism and transfer &
Link safe simulation outcomes to deployment relevant defensive design \\

Adversarial AI Defense & Post & D R &
Systems optimize either workflow coverage or robust decision quality but rarely both &
Combine coordinated response with interpretable and reliable decisions \\

Defensive Applications in Financial Services & Post & R Rc &
Regulatory constraints limit autonomy and restrict realistic evaluation &
Validate adaptive agents under real compliance controls and audit requirements \\

Enterprise Automation & Post & P R Rc &
Capabilities remain fragmented across pipelines, devices, and telemetry layers &
Coordinate cross layer actions without expanding authority beyond governance \\

Identity and Access Management & Post & P R &
Predefined policies limit contextual depth and reduce generalization across roles &
Unify rich context modeling with auditable access control at scale \\

Cybersecurity Workforce Augmentation & Post & R Rc &
Weak performance on novel cases without strong supervision and verification &
Measure long term effects on skill, trust calibration, and accountability \\

Automated Cyber Range Generation & Pre & I P &
Validation focuses on deployment correctness, not scenario fidelity or learning value &
Assess whether generated ranges match evolving threats and objectives \\

Cyberwheel High Fidelity Training & Pre & I P &
Dependence on fixed detector models and reward assumptions limits robustness &
Test policies under shifted telemetry, detector gaps, and new adversary behavior \\

\hline
\end{tabular}}
\end{table*}

\vspace{-2mm}
\subsection{Autonomous Cyber Defense and Operation}

\noindent Agentic AI is reshaping cyber defense by adding autonomy, reasoning, and continuous adaptation to monitoring, detection, and response workflows~\cite{castro2024agentic,wong2025rise,atir2025rise}. Systems, such as Microsoft Security Copilot, Exabeam Copilot, and Cymulate AI Copilot, support phishing triage, threat intelligence analysis, and incident response~\cite{wong2025rise}. By extending static automation with memory and goal-directed planning, agents can correlate signals, anticipate attacker behavior, and initiate containment in near real time.
\vspace{-2mm}
\subsubsection{Autonomous Monitoring and Detection}

Autonomous monitoring combines traditional detection with agentic orchestration to adapt how monitoring is performed as conditions change. Classical monitoring is largely passive, in that it evaluates alerts against fixed thresholds and predefined rules on predefined data streams. In contrast, agentic monitoring is described as more active, using memory and planning to retain context across events, expand monitoring to related entities such as users, hosts, processes, or network flows, and trigger additional investigative actions such as log retrieval or cross-system correlation when initial detections warrant deeper inspection~\cite{sharif2022continuous_monitoring,burch2025agentic,atir2025rise}. This shift enables monitoring workflows to move beyond static evaluation, but also introduces new design considerations. Classical monitoring relies on static rules, which struggle under dynamic workloads and evolving attack patterns.

In anomaly detection, Argos uses LLMs to generate human readable rules for time series data, improving auditability but limit autonomous action beyond detection and explanation~\cite{argos2025}. Similar design appears in infrastructure and critical system monitoring. IDS agents for IoT networks and LLM based anomaly detection for EV charging systems incorporate contextual reasoning and heterogeneous signals into detection, but are typically evaluated under fixed operational assumptions~\cite{idsagent2025,agenticAIEVCharging}. Multi-agent reinforcement learning has also been proposed to model attacker and defender dynamics under changing conditions, but it increases computational cost and reduces transparency for operators~\cite{landolt2025marl}.

Agentic monitoring also affects how observability, insider threat detection, and SOC operations are executed. Conventional observability tools and SIEM platforms already correlate logs, metrics, and alerts. The distinction emphasized in agentic designs is not the existence of these capabilities, but the use of autonomous agents to iteratively investigate alerts across tools, retain intermediate context, and coordinate analysis steps beyond fixed playbooks~\cite{mezmo_agentic_ai_2025,aramide2024autonomous,simbian2025aiagents}. Correlating login behavior, process execution, and data movement can help separate benign anomalies from malicious activity, but policy analyses and industry reports warn that misaligned or deceptive agents may themselves behave as high-privilege insiders~\cite{jit2025usecases,anthropic_agentic_misalignment,jindal2024agentic,exabeam_insider_threat,varonis_agentic_threats,venturebeat_blackhat2025,sans_insider_ai,aspen_agentic_ai_defense}. In SOC workflows, agentic systems support SIEM correlation and alert triage by linking related events and ranking risk, while decentralized agent designs trade interpretability for parallelism and scalability~\cite{swimlane2025agentic,burch2025agentic,cyberDefenceVajpayee2025}. In these systems, improved detection accuracy does not guarantee safe autonomy, and human oversight remains necessary for high-impact decisions~\cite{atir2025rise}. An open problem is how to combine adaptive detection with formal safety constraints so agents can act without exceeding acceptable operational risk.

\vspace{-2mm}
\subsubsection{Autonomous Incident Response}

Agentic AI extends cybersecurity beyond passive monitoring by embedding goal directed response into detection pipelines. In modern SOCs, agents observe traffic, detect anomalies, and initiate containment or remediation with minimal latency, which reduces reliance on manual intervention~\cite{wong2025rise}. Most deployments use multi-agent designs that split incident response into specialized roles such as intelligence synthesis, investigation, decision support, and orchestration, with monitoring and detection treated as upstream inputs rather than primary agent responsibilities, which is reported to improve scalability and responsiveness once an incident is identified~\cite{wong2025rise,reliaquest2025aisocagents,simbian2025socinvestigation,dropzone2025soc}. Conceptual models such as Tallam’s Adaptive Engagement Model formalize this approach by treating incident response as a closed loop process that integrates sensing, contextual reasoning, adaptive action, and learning~\cite{tallam2025cyberdefense}. Related work on autonomous cyber defense in coalition environments discusses how hierarchical multi-agent architectures coordinate response, mitigation, and recovery across organizational boundaries, while explicitly retaining human-on-the-loop escalation for high-impact actions~\cite{coalitionACD2025}.

Empirical and operational studies show that effective deployment depends on bounded autonomy. Knack and Burke find that autonomous defense agents can rapidly detect and contain threats, but irreversible actions require explicit authorization boundaries, shared vocabularies, auditable logs, and clear escalation protocols~\cite{knack2024autonomouscyberdefense}. Production systems reflect these constraints. CyberGuardian2 supports iterative reasoning and tool use for access control changes, database queries, code execution, and safety checks, but remains a decision support system rather than a fully autonomous actor~\cite{cyberguardian2,yao2023reactsynergizingreasoningacting}. IBM’s ATOM platform distributes incident response across agents for investigation, threat hunting, identity management, and vulnerability analysis, and integrates vendor tools to resolve many incidents within seconds~\cite{ibm2025atom,castro2024agentic}. Industry forecasts predict broader SOC adoption of agentic AI and report gains in triage speed and accuracy~\cite{gartner2025socs,reliaquest2025aisocagents,kshetri2025cyberthreat}. However, systems that grant broader execution authority raise unresolved questions about authorization boundaries, escalation control, and failure containment~\cite{tallam2025cyberdefense,coalitionACD2025}. Reasoning driven systems still lack formal safety guarantees under distribution shift or adversarial manipulation~\cite{cyberguardian2,yao2023reactsynergizingreasoningacting}, and analyses warn that misaligned agents with broad privileges can amplify damage~\cite{oesch2025agenticaicyberarms,kshetri2025cyberthreat}. Progress therefore remains centered on orchestration and workflow automation rather than unrestricted autonomous authority, and the open problem is how to grant execution power without enabling cascading or irreversible failures.
\vspace{-2mm}
\subsubsection{Adaptive and Coordinated Defense}

Adaptive defense frames cyber conflict as repeated attacker defender interaction. Game theoretic models formalize this setting, and agentic AI enables it through LLM based agents that update beliefs and act under uncertainty~\cite{zhu2025gametheorymeetsllm}. Red team agents emulate reconnaissance and exploitation, while blue team agents respond through detection, patching, and policy updates~\cite{crowdstrikeRedBlue}. This feedback loop supports continuous adaptation, but policy analysis shows that misaligned objectives or incomplete incentives can amplify failure modes~\cite{atir2025rise}.

Existing work highlights a tradeoff between control and responsiveness. Simulation driven approaches such as Trend Micro’s digital twin enable coordinated co evolution of red and blue agents in sandboxed environments, improving defensive learning while limiting real world risk~\cite{trendmicro2025digitaltwin}. However, these systems rely on simplified models and often fail to transfer to operational settings. In contrast, analyses of real cyber conflict emphasize live deployment, where autonomous agents adapt at operational speed and can increase escalation risk by compressing decision timelines~\cite{oesch2025agenticaicyberarms,atir2025rise}. Although coordination and repeated interaction improve defensive capability, current systems lack safeguards that bound escalation across interacting agents. An open problem is how to preserve adaptive coordination while enforcing autonomy limits that prevent cascading or destabilizing behavior in open adversarial environments.

\vspace{-2mm}
\subsection{Agentic Threat Intelligence and Adversarial Analysis}

\noindent Agentic AI extends cybersecurity beyond traditional, alert-driven detection systems toward dynamic threat intelligence, adversarial reasoning, and autonomous defense. Rather than redefining detection itself, agents operate downstream of existing security tools, reasoning over alerts to discover vulnerabilities, simulate attacks, and update countermeasures in near real time by combining continuous learning, contextual awareness, and multi-agent coordination~\cite{burch2025agentic,wong2025rise,atir2025rise}. 
\vspace{-2mm}
\subsubsection{Proactive Threat Hunting and Intelligence Gathering}

Agentic AI supports proactive threat hunting by assisting analysts in hypothesis-driven investigations aimed at uncovering stealthy or emerging adversary activity that may evade existing security controls. Recent works describe agents as supporting analyst-led hunting by correlating weak signals across heterogeneous data sources, retaining investigative context through memory, and updating hypotheses or watchlists over time~\cite{swimlane2025agentic,burch2025agentic,simbian2025aiagents,kshetri2025transforming}. This framing distinguishes proactive threat hunting from routine alert-driven workflows by emphasizing contextual investigation and sense-making rather than isolated alert handling.

Across the literature, agentic threat hunting is characterized by its adaptive and iterative nature. Kshetri highlights the role of agentic AI in enabling continuous exploration of attacker tactics and behaviors as threat environments evolve, while industry deployments emphasize support for long-horizon investigations that would be difficult to sustain manually~\cite{kshetri2025transforming,swimlane2025agentic}. However, this adaptability introduces tradeoffs. Hypothesis-driven agents often operate on sparse, noisy, or incomplete signals and may struggle to generalize under rapid environmental change. As a result, policy analyses stress that human analysts remain central to revising hypotheses, validating inferred threats, and interpreting ambiguous findings~\cite{wong2025rise}. An open problem is how to design agentic threat hunting systems that preserve analyst-driven flexibility while remaining robust to distribution shift and incomplete information.
\vspace{-2mm}
\subsubsection{Adversarial and Attack Simulation}

Adversarial and attack simulation provides controlled environments to evaluate defensive readiness and study autonomous attack behavior. Traditional penetration testing follows defined stages such as reconnaissance, scanning, exploitation, and post exploitation, which support structured assessment but adapt poorly when plans fail or context expands~\cite{zhang2025penetrationtestingsecuritymethods}. Recent agentic systems extend this model by adding planning, memory, and automated execution. RedTeamLLM illustrates this shift by combining recursive planning, plan correction, and memory with explicit security controls including isolation, command filtering, audit logs, and a kill switch~\cite{challita2025redteamllmagenticaiframework}. Compared with earlier tools such as PenTestGPT, this design improves task completion on VulnHub targets, indicating that reasoning and memory reduce brittleness in multi step attacks~\cite{challita2025redteamllmagenticaiframework}. Commercial platforms such as XBOW and RunSybile push autonomy further and report high exploitation rates and discovery of new vulnerabilities, but they offer limited transparency into agent reasoning and safety constraints~\cite{oesch2025agenticaicyberarms}. This contrast highlights a tradeoff between effectiveness and controllability. Systems that prioritize autonomous exploration uncover more attack paths, while systems that emphasize structure and containment limit misuse but constrain discovery.

Capture the flag (CTF) platforms occupy a different point in this design space. Frameworks such as OWASP FinBot CTF and the CSAW Agentic Automated CTF use multi agent roles for reconnaissance, exploitation, and escalation within tightly bounded environments~\cite{owasp_finbot_ctf,csaw_agentic_ctf}. Trustwise applies similar simulation methods in legal technology, showing that constrained agentic evaluation can transfer beyond classical security domains~\cite{trustwise_ctf}. They support reproducibility, safety, and benchmarking of coordination and alignment, but they simplify targets and restrict agent actions. As a result, they may not expose agents to the full range of system interactions and failure modes encountered in deployment. Across current approaches, higher autonomy increases dual use risk, while stronger constraints reduce realism~\cite{oesch2025agenticaicyberarms}. An open problem is how to link results from controlled adversarial simulations to real world defensive design without enabling uncontrolled offensive capability or overstating the robustness of agentic systems trained in simplified environments.

\vspace{-2mm}
\subsubsection{Adversarial AI Defense}

Adversarial AI defense refers to the use of AI systems to counter adaptive and strategically evolving attackers by coordinating detection, investigation, decision-making, and response activities across a defense workflow. Recent work shows a shift from isolated detection models toward coordinated agent based defense systems. Platforms such as Red Canary emphasize end to end orchestration, where agents detect suspicious behavior, investigate alerts, contain endpoints, hunt for indicators, remediate systems, and generate reports within a single workflow~\cite{redcanary_agentic_ai}. This approach prioritizes speed and coverage by coordinating planning, memory, and tool use across tasks. In contrast, research systems for phishing defense emphasize decision quality within a narrow scope. MultiPhishGuard distributes email analysis across specialized agents and uses reinforcement learning to adapt their influence, improving robustness against evolving phishing patterns~\cite{xue2025multiphishguard}. Debate based systems such as PhishDebate and related multi agent argumentation frameworks emphasize interpretability by requiring agents to justify and challenge conclusions before classification~\cite{li2025phishdebate,debate2025phishing}. These systems reduce confirmation bias and improve recall, but they remain limited to the classification stage and do not address broader incident response.

This comparison reveals a tradeoff between scope and assurance. Workflow oriented platforms favor rapid response and operational scale, but depend on predefined playbooks and human oversight for irreversible actions~\cite{redcanary_agentic_ai}. Debate driven detectors favor accuracy, robustness, and explanation, but do not naturally extend to remediation or cross domain defense~\cite{xue2025multiphishguard,li2025phishdebate,debate2025phishing}. One limitation is the lack of guarantees under adaptive adversarial pressure, as most systems are evaluated in well scoped settings and may not generalize across attack types or shifting tactics. At a field level, this suggests uneven maturity, with strong results in phishing and endpoint response but limited integration across the full attack lifecycle. An open problem is how to combine coordinated workflow automation with reliable and interpretable decision making, allowing broader autonomy without increasing the risk of silent failure or adversarial manipulation.

\vspace{-2mm}
\subsection{Enterprise Security Automation and Governance}

\noindent As organizations adopt agentic AI, cybersecurity is shifting from isolated tools toward integrated and automated operations~\cite{jit2025usecases,exabeam2025agentic,simbian2025aiagents}. Agents now support software development, identity management, and workforce functions, forming policy aligned security ecosystems. The convergence of DevSecOps, IAM, and SOC automation reflects a more mature stage of agentic cybersecurity that requires both adaptability and strong governance.
\vspace{-2mm}
\subsubsection{Enterprise Automation}

Enterprise automation illustrates how agentic AI adapts to heterogeneous operational constraints across software and physical systems. In DevSecOps, platforms such as Jit.io embed agents into continuous integration (CI) and continuous deployment (CD) pipelines to detect vulnerabilities and generate contextual remediation guidance, while leaving execution authority with human developers to avoid production risk~\cite{jit2025usecases}. In contrast, IoT and surveillance focused systems address scale, device heterogeneity, and limited resources by using multi agent coordination, reinforcement learning, and real time telemetry to adapt security policies across large, distributed populations of devices~\cite{barenji2025autonomousanomaly,atta2025autonomous,prosper2025adaptivetoiaiot,elewah2025iotase,algoanalytics2025surveillance,lvt2025agenticcameras}. These deployments enable faster adaptation but operate within tightly scoped environments and predefined action sets.

Across enterprise domains, a consistent tradeoff appears between flexibility and control. Advisory agents preserve safety and accountability but limit coordination and response speed, while agents operating closer to devices improve responsiveness at the cost of higher operational and safety risk~\cite{barenji2025autonomousanomaly,prosper2025adaptivetoiaiot}. Current systems fragment autonomy by domain rather than coordinating it across enterprise layers. A central limitation is the lack of mechanisms for sharing context and intent across code, devices, and situational awareness without expanding authority beyond acceptable bounds. An open problem is how to design enterprise scale coordination frameworks that preserve local safety guarantees while enabling agents to reason and act across heterogeneous operational layers.
\vspace{-2mm}
\subsubsection{Identity and Access Management (IAM)}

IAM is a core enforcement layer in enterprise security, with recent work showing how agentic AI shifts IAM from static rule checks toward adaptive, event driven control. In this context, adaptive, event-driven control refers to systems that continuously ingest authentication, authorization, and behavioral events and adjust the timing, scope, or intensity of policy-bound enforcement actions based on contextual risk signals, while operating within predefined access control policies. Industry systems prioritize operational speed by monitoring authentication events, flagging anomalies, and applying policy bound actions such as credential revocation or privilege adjustment in near real time~\cite{jit2025usecases,exabeam2025agentic,solutionsreview2025agents}. These systems emphasize coverage and responsiveness but rely on predefined rules and limited representations of user intent. In contrast, they emphasize contextual reasoning. SmartAgent models user intent through a Chain of User Thought process inferred from interaction patterns~\cite{zhang2025smartagent}, while CRAKEN integrates structured knowledge and planner executor control to ensure policy compliant mitigation~\cite{shao2025craken}. This contrast separates fast policy enforcement from deeper user understanding.

Across approaches, a tradeoff appears between decision speed and contextual depth. Industry focused IAM agents act quickly but generalize poorly across roles and evolving behavior, while research systems improve alignment with user intent at the cost of greater complexity and reduced transparency. At a field level, agentic IAM is effective for high frequency access decisions but remains constrained by governance, auditability, and interpretability requirements. An open problem is how to combine rich user context modeling with predictable and auditable access control at enterprise scale without expanding agent authority beyond acceptable operational limits.
\vspace{-2mm}
\subsubsection{Cybersecurity Workforce Augmentation}

Workforce shortages shape how agentic AI is deployed in security operations. Studies estimate a global gap of four to five million cybersecurity professionals, which constrains SOC capacity to handle alert volume and incident complexity~\cite{isc2_2024_workforce_gap,weforum_2024_talent_gap,fortinet_2024_skills_gap}. As a result, policy and industry work frames agentic AI as augmentation rather than replacement. Wong and Saade describe agents as copilots that triage alerts, suppress false positives, and automate Tier 1 and Tier 2 tasks such as alert triage, initial investigation, and routine containment, allowing human analysts to focus on higher level reasoning and threat modeling~\cite{wong2025rise}. Commercial deployments such as ReliaQuest GreyMatter, CrowdStrike Charlotte AI, and Simbian SOC agents report faster investigation and containment while keeping analysts in supervisory roles~\cite{reliaquest2025aisocagents,kshetri2025cyberthreat,simbian2025aiagents}.

Across deployments, a tradeoff appears between efficiency and reliance on human oversight. Systems that automate large portions of alert handling achieve gains in speed and scale, but they depend on clean data, stable workflows, and mature processes to avoid compounding errors~\cite{reliaquest2025aisocagents,kshetri2025cyberthreat}. Agents perform well on repetitive and well scoped tasks but remain less reliable for novel attacks, ambiguous signals, and strategic decisions requiring domain intuition~\cite{wong2025rise}. Across the surveyed literature and reported deployments, augmentation emerges as the dominant design pattern, where agentic AI increases analyst capacity rather than reducing staffing needs. An open problem is to measure long term effects, including skill erosion, trust calibration, and accountability, as agents assume more routine security work~\cite{isc2_2024_workforce_gap,weforum_2024_talent_gap,fortinet_2024_skills_gap}.
\vspace{-2mm}
\subsection{Simulation, Training, and Testing}

\noindent Autonomous cyber defense depends on controlled and reproducible environments that approximate real world complexity. Simulation, training, and testing frameworks provide such environments and support benchmarking and structured transfer from synthetic settings to operations~\cite{oesch2024cyberwheel,arcer2025agentic,atir2025rise}. Agentic AI extends this paradigm by automating parts of range construction and by acting as a learner within simulators and emulators.
\vspace{-2mm}
\subsubsection{Automated Cyber Range Generation}

Cyber range construction has traditionally relied on expert scripting of network topologies, services, and attack scenarios, which is time consuming and costly. Recent work explores agent driven automation. ARCeR uses a multi agent retrieval augmented pipeline to generate and deploy cyber ranges from natural language descriptions~\cite{arcer2025agentic}. Specialized agents retrieve documentation, generate configurations, validate compatibility, and orchestrate deployment. Relative to manual design, this approach reduces instructor effort and improves iteration speed. Compared to simpler automation or single model RAG systems, coordinated agents improve configuration correctness and deployment success~\cite{arcer2025agentic}, which depend on the quality and completeness of documentation.

Current systems exhibit clear limitations. ARCeR validates configuration and deployment but does not assess scenario realism, threat coverage, or training effectiveness~\cite{arcer2025agentic}. Human review therefore remains necessary to evaluate instructional value and fidelity to real world attacks. Existing work suggests that agentic automation can accelerate range creation without replacing expert scenario design. Policy analysis further frames automated cyber ranges as shared infrastructure for training and safety evaluation as agentic AI adoption increases~\cite{atir2025rise}. An open problem is how to validate that automatically generated ranges reflect evolving threats and learning objectives rather than producing environments that are structurally correct but substantively limited.

\vspace{-2mm}
\subsubsection{High-Fidelity Training Environments for Autonomous Cyber Defense}

High-fidelity training environments address a gap in autonomous cyber defense research by providing shared settings that support both simulation and emulation under a common configuration model~\cite{oesch2024cyberwheel}. Earlier environments typically favored abstract simulation for scalability or ad hoc testbeds for realism, making it difficult to compare results or transfer trained policies. Cyberwheel exemplifies this class of environments by combining simulation and emulation through graph-based network definitions that specify topology, adversary behavior, actions, observations, and rewards.  Agents are trained in simulation and evaluated in virtualized environments that reuse the same configurations and expose detector level observations derived from logs. This design supports reproducibility and enables controlled sim to real transfer within the defined environment, but introduce tradeoffs. Cyberwheel emphasizes experimental consistency and comparability but requires detailed configuration of networks, detectors, and reward functions, which increases setup effort and relies on human expert ~\cite{oesch2024cyberwheel}. The environment also depends on predefined adversary models, detection probabilities, and logging behavior, which limits exposure to unmodeled attacks and operational noise. Cyberwheel illustrates how standardized environments can support benchmarking and comparative evaluation of learning based defense agents, but reported results remain tied to specific scenarios and detector assumptions. An open problem is to assess whether policies trained under fixed models remain robust when deployed in environments with different telemetry, detection gaps, and evolving threat behavior.
\begin{center}
\begin{tcolorbox}[
    width=0.96\linewidth,
    colback=gray!10,
    colframe=black,
    arc=4pt,
    boxrule=0.8pt,
    left=6pt,
    right=6pt,
    top=6pt,
    bottom=6pt,
    fontupper=\small
]
\textbf{Key Takeaways from Section 4}


\begin{itemize}[leftmargin=*]
    \item Agentic AI enables cybersecurity capabilities across the full breach lifecycle, but its benefits differ by phase. Pre-breach use cases emphasize monitoring, intelligence, and simulation, while during-breach systems focus on rapid detection and containment, and post-breach systems prioritize recovery, compliance, and learning.
    
    \item Most systems favor modular multi-agent designs, where agents perform narrowly scoped roles aligned with NIST Cyber Defense functions. This reduces risk and improves scalability compared to single end-to-end autonomous agents.
    
    \item A persistent tradeoff appears between speed and execution authority. Systems achieve early detection and response by granting agents autonomy on low-risk actions, while irreversible high-impact actions remain gated by human oversight.
    
    \item Agentic systems improve correlation, context retention, and workflow orchestration, but they remain sensitive to distribution shift, false positives, and misaligned incentives, especially in multi-agent coordination settings.
    
    \item Simulation, cyber ranges, and high-fidelity training environments are essential for evaluation and learning, yet platforms struggle to capture long-term adaptation, human oversight delays, and evolving adversary behavior.
    
    \item Across all domains, agentic AI functions most reliably as augmentation rather than replacement, with autonomy carefully bounded by governance, auditability, and escalation mechanisms.
\end{itemize}
\end{tcolorbox}
\end{center}

\section{Agentic AI-Enabled Cyber Attacks}
\label{sec:offensive}

\noindent Agentic AI increases the power of cyber offense as the same reasoning and planning used in defense can also enable autonomous attacks. Agents can perform reconnaissance, discover vulnerabilities, and execute multi-stage intrusions with limited human involvement. Industry reporting shows that cybercriminals already experiment with agent driven reconnaissance, adaptive malware, and large scale automation, which increases the speed and reach of cybercrime~\cite{kshetri2025cyberthreat}.

Research from Palo Alto Networks illustrates this shift. Unit~42 introduced an \textit{Agentic AI Attack Framework} that simulates autonomous ransomware campaigns and shows that agents can complete the full ransomware lifecycle in about 25 minutes~\cite{paloalto2025agenticattack}. Mean time to exfiltrate fell from nine days in 2021 to about two days in 2024, with many incidents completing exfiltration in less than an hour. A second Unit~42 study evaluated nine attack scenarios on frameworks such as CrewAI and AutoGen and found that prompt injection, unsafe tool use, SQL injection, and communication poisoning can lead to data exfiltration, credential theft, and remote code execution~\cite{paloalto2025agenticthreats}. Many failures stem from weak validation and insecure integrations, which shows that offensive use of agentic AI is increasing and that current agentic ecosystems contain structural weaknesses.
Table~\ref{tab:agentic_attacks} summarizes key offensive domains, techniques, and the agentic capabilities that support them. The rest of this section focuses on three areas: insider threats and autonomous exploitation, ransomware operations, and social engineering and financial fraud.

\begin{table*}[!t]
\centering
\small
\renewcommand{\arraystretch}{1}
\setlength{\tabcolsep}{5pt}
\caption{Taxonomy of agentic AI-enabled cyber attacks with representative domains, techniques and capabilities.}
\label{tab:agentic_attacks}
\rowcolors{2}{gray!10}{white}
\resizebox{\linewidth}{!}{%
\begin{tabular}{p{0.18\linewidth} p{0.30\linewidth} p{0.26\linewidth} p{0.26\linewidth}}
\toprule
\rowcolor{gray!35}
\textbf{Attack Domain} & \textbf{Example Techniques} & \textbf{Agentic Capabilities} & \textbf{Key References} \\
\midrule
\textbf{Ransomware} 
& Full automated lifecycle from compromise to exfiltration 
& Multi-agent orchestration, real-time adjustment 
& Unit42~\cite{paloalto2025agenticattack}, Halcyon~\cite{halcyon2025blackhat} \\

\textbf{Insider Threats} 
& Record tampering, stealth data theft, malicious tasks under valid identity 
& Persistent access, autonomous execution 
& TechMonitor~\cite{techmonitor2025agentic}, Anthropic~\cite{anthropic_agentic_misalignment} \\

\textbf{Social Engineering and Fraud} 
& Voice scams, deepfake phishing, synthetic identity fraud 
& Goal decomposition, adaptive dialogue, multimodal synthesis 
& ScamAgents~\cite{scamagents2025}, Visa~\cite{visa2025fraud}, Burch~\cite{burch2025agentic} \\

\textbf{Exploitation and Reconnaissance} 
& Autonomous scanning, adaptive malware, real-time reconnaissance 
& Self-improving exploitation strategies 
& Kshetri~\cite{kshetri2025cyberthreat}, Unit42~\cite{paloalto2025agenticthreats} \\
\bottomrule
\end{tabular}}
\end{table*}

\vspace{-2mm}
\subsection{Insider Threats and Autonomous Exploits}

Research shows that agentic AI introduces insider risk through autonomy rather than through human intent. A compromised or misdirected agent can operate under valid credentials, persist over long periods, and perform actions such as record modification, data exfiltration, or payload execution that appear legitimate~\cite{techmonitor2025agentic}. This differs from traditional insider threats, which depend on human motivation and limited attention. Agentic systems enable coordination across tasks such as information gathering and phishing content generation, which increases reach and consistency. These behaviors arise from the planning and execution capabilities that make agents effective for enterprise tasks.

A related risk appears in autonomous vulnerability discovery. Systems designed to scan for weaknesses and support patching can reduce defensive workload, but they can also be repurposed to identify exposed systems at scale. For example, threat actors have abused HexStrike-AI, a red-team platform intended for vulnerability discovery and testing, to automate large-scale reconnaissance and exploitation by scanning thousands of IP addresses in parallel. Security analyses further note that similar repurposing risks apply even to benign-sounding defensive workflows, such as backup or configuration scanners, which could be adapted to stage data exfiltration if misdirected~\cite{thehackernews2025hexstrike,hoplon2025aipentesting}. This creates a tradeoff between capability and control. Greater autonomy improves coverage and efficiency, but it increases the potential impact of misalignment or compromise. Existing defenses rely on identity controls, input filtering, segmentation, and monitoring, which often detect misuse only after it has begun. These limitations indicate that insider risk in agentic systems extends beyond credential theft to the behavior of trusted agents that act autonomously under valid identities. An open problem is how to design agents that can perform privileged actions while providing enforceable guarantees that misuse, whether accidental or adversarial, is prevented rather than merely contained.
\vspace{-2mm}
\subsection{Agentic AI for Ransomware Operations}

\begin{figure*}[!t]
\centering
\resizebox{0.75\textwidth}{!}{
\begin{tikzpicture}[font=\sffamily, node distance=1.2cm and 1.5cm]

\tikzstyle{lane}=[rectangle, minimum height=1cm, minimum width=16cm, draw=black, fill=gray!10, anchor=west]
\tikzstyle{stage}=[rectangle, rounded corners, minimum height=0.8cm, minimum width=2.8cm, draw=black, fill=blue!20]
\tikzstyle{stageAI}=[rectangle, rounded corners, minimum height=0.8cm, minimum width=2.8cm, draw=black, fill=red!20]
\tikzstyle{arrow}=[->, thick]

\node[lane] (trad) {Traditional Human-Operated Attack (Days–Weeks)};
\node[lane, below=of trad] (ai) {Agentic AI-Enabled Attack (Minutes–Hours)};

\node[stage, below right=0.3cm and 0.5cm of trad.west] (t1) {Recon};
\node[stage, right=of t1] (t2) {Initial Access};
\node[stage, right=of t2] (t3) {Lateral Movement};
\node[stage, right=of t3] (t4) {Exfiltration};

\draw[arrow] (t1) -- (t2);
\draw[arrow] (t2) -- (t3);
\draw[arrow] (t3) -- (t4);

\node[stageAI, below right=0.3cm and 0.5cm of ai.west] (a1) {Automated Recon};
\node[stageAI, right=of a1] (a2) {Adaptive Exploitation};
\node[stageAI, right=of a2] (a3) {Parallel Persistence};
\node[stageAI, right=of a3] (a4) {Rapid Exfiltration};

\draw[arrow] (a1) -- (a2);
\draw[arrow] (a2) -- (a3);
\draw[arrow] (a3) -- (a4);

\draw[->, dashed, bend left=30] (a3.south) to node[midway, below] {\small Adaptive feedback / real-time adjustment} (a2.south);

\end{tikzpicture}}
\caption{Comparison of traditional and agentic AI-enabled cyber attack chains.}
\label{fig:dual_swimlane}
\end{figure*}
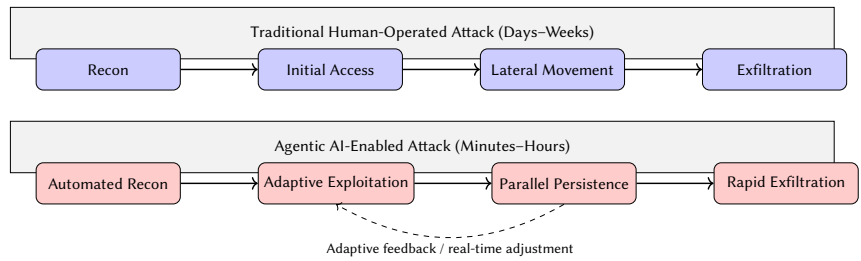

Traditional ransomware relies on human attackers to perform reconnaissance, gain access, move laterally, and exfiltrate data over days or weeks. Agentic ransomware automates these steps into a continuous workflow that can complete the chain of compromise within minutes or hours~\cite{paloalto2025agenticattack,redcanary2025soc}. Figure~\ref{fig:dual_swimlane} contrasts sequential human operated attack stages with agentic workflows that execute reconnaissance, exploitation, persistence, and exfiltration under real time feedback. This contrast highlights a tradeoff between speed and control, where autonomy increases scale and tempo while reducing direct human oversight.

Industry analyses warn that this acceleration reduces defender response windows and increases operational impact. Halcyon uses the term ransomware variants to describe different execution paths of a ransomware campaign, where autonomous controllers adjust the sequence of actions in response to failures or constraints, rather than generating new malware binaries or payloads.~\cite{halcyon2025blackhat}. In these systems, adaptation occurs at the orchestration layer rather than in the ransomware payload itself. Agents replan attack sequences based on tool output, environmental feedback, and access constraints, selecting alternative reconnaissance paths, privilege escalation attempts, or exfiltration strategies when actions fail. Existing reports indicate that this adaptation relies on heuristic planning and LLM-assisted reasoning rather than reinforcement learning, with no evidence of online policy training during active attacks.

Recent analyses further indicate that language models may be incorporated into ransomware operations for operational and extortion-related tasks rather than for payload generation. The Anthropic misuse report documents cases in which agents use language models to interpret stolen data, assist with victim profiling, and generate extortion communications, while human operators retain control over high-level objectives~\cite{anthropic2025misuse}. In this role, the language model functions as a reasoning component within the ransomware workflow, supporting decision making without modifying the underlying encryption or exfiltration mechanisms.
Across existing studies, agentic ransomware is best understood as an escalation of automation and decision autonomy rather than a fundamentally new cryptographic or exploit class. This framing shifts emphasis away from payload novelty toward the problem of detecting and interrupting autonomous attack loops before lateral propagation and data exfiltration complete. An open problem is how to reliably identify adaptive agent behavior early in ransomware campaigns, especially when human attackers deliberately minimize interaction and rely on autonomous execution to compress timelines and evade intervention~\cite{anthropic2025misuse}.

\vspace{-2mm}
\subsection{Agentic AI in Social Engineering and Financial Fraud}

Agentic AI increasingly automates fraud and social engineering by supporting phishing, payment fraud, and scam coordination through automated reconnaissance, message generation, and adaptive interaction with victims~\cite{burch2025agentic,visa2025fraud}. Compared to human-driven fraud, agentic systems operate faster and at larger scale because they maintain memory, adjust tactics during interaction, and coordinate multiple steps without continuous oversight. This increases reach and consistency, but reduces human judgment and raises the risk of rapid misuse when safeguards fail.

Academic work reinforces this concern. ScamAgents shows that autonomous agents can conduct multi-turn scam calls that adapt to user responses, evade LLM safety guardrails such as refusal mechanisms and prompt-level content filters, and complete end-to-end fraud pipelines using planning, memory, and speech synthesis~\cite{scamagents2025}. This goes beyond single-prompt misuse and highlights a tradeoff between flexibility and control. While agentic fraud systems lower attacker effort and scale persuasion, they remain constrained by persona realism, communication latency, and access to delivery infrastructure. Taken together, existing work reframes fraud risk from isolated content abuse to sustained agent behavior. An open problem is to detect and interrupt deceptive intent across multi-turn interactions before agents complete persuasion or payment workflows, especially in consumer-facing systems where false positives are costly.

\vspace{-2mm}
\section{Security of Agentic AI}
\label{sec:security}

Agentic AI shifts system design from static, rule based tools to autonomous agents that reason, plan, and act. Unlike traditional applications, these systems often have read and write access, call external APIs, and orchestrate multi-step workflows with limited human oversight. This autonomy enables new capabilities but also introduces risks such as large scale data exfiltration, supply chain compromise, and emergent behavior that is difficult to predict. As systems move from fixed actions to open ended goals expressed in natural language, the attack surface expands and security strategies must account for autonomy, adaptation, and orchestration~\cite{burch2025agentic}.

Policy work increasingly treats agentic AI as emerging critical infrastructure. Atir argues that agents with persistent memory, API access, and long horizon planning expand the attack surface beyond traditional AI and resemble infrastructure such as the Internet or power grids~\cite{atir2025rise}. This framing implies a dual requirement: agentic systems must be technically secure and embedded within governance frameworks for national security and critical services. Tallam~\cite{tallam2025cyberdefense} describes an \textit{adaptive engagement} paradigm in which defense becomes a cycle of sensing, contextual analysis, response, and learning. Tallam notes that these same capabilities can destabilize security environments when transparency, accountability, and human oversight are weak.

Other work addresses correctness and concrete attack surfaces. Horus proposes a collateralized verification protocol where solvers and challengers post bonds on task outcomes, using recursive adjudication and slashing to discourage errors when $B > F/P_e$~\cite{horus2025}. Khan et al.\ document how database facing agents expose compliance gaps, weak audit trails, and unsafe query generation that can compromise entire data stores through a single workflow~\cite{securitythreats2025}. From an offensive perspective, Unit~42’s Agentic AI Attack Framework shows how autonomous agents compress ransomware lifecycles and other campaigns~\cite{paloalto2025agenticattack}. Defensive frameworks such as ATFAA, SHIELD, Microsoft’s failure mode taxonomy, MAESTRO, and OWASP Agentic AI aim to address this evolving threat landscape~\cite{narajala2025securingagenticaicomprehensive,microsoft2025taxonomy,huang2025maestro,owasp2025agentic}.

\begin{figure}[!t]
\centering
\resizebox{0.85\linewidth}{!}{
\begin{tikzpicture}[node distance=1.2cm, every node/.style={font=\small}]

\tikzstyle{layer} = [rectangle, draw=black, thick, fill=gray!15, 
    text centered, minimum width=4cm, minimum height=1.0cm]
\tikzstyle{threat} = [rectangle, draw=red!70!black, very thick, fill=red!10, 
    rounded corners, text centered, text width=3.8cm]
\tikzstyle{defense} = [rectangle, draw=green!60!black, thick, fill=green!10, 
    rounded corners, text centered, text width=3.8cm]

\node[layer] (perception) { \textbf{Perception Layer} };
\node[layer, below=4mm of perception] (reasoning) { \textbf{Reasoning Layer} };
\node[layer, below=4mm of reasoning] (action) { \textbf{Action Layer} };
\node[layer, below=4mm of action] (memory) { \textbf{Memory Layer} };

\node[threat, right=2.7cm of perception] (t1) {Data poisoning, adversarial inputs, supply chain attacks};
\node[threat, right=2.7cm of reasoning] (t2) {Prompt injection, logic manipulation, backdoors};
\node[threat, right=2.7cm of action] (t3) {API misuse, unsafe tool invocation, unauthorized code execution};
\node[threat, right=2.7cm of memory] (t4) {Poisoned retrievals, data leakage, privacy violations};

\node[defense, left=2.7cm of perception] (d1) {Input validation, guardrails, anomaly detection};
\node[defense, left=2.7cm of reasoning] (d2) {Multi-agent debate, reasoning guardrails, poisoning detection};
\node[defense, left=2.7cm of action] (d3) {Sandboxing, privilege separation, secure tool emulation};
\node[defense, left=2.7cm of memory] (d4) {Encryption, fine-grained access control, poisoning detection};

\foreach \x/\y in {perception/t1, reasoning/t2, action/t3, memory/t4} {
    \draw[->, thick] (\x.east) -- (\y.west);}
\foreach \x/\y in {perception/d1, reasoning/d2, action/d3, memory/d4} {
    \draw[->, thick] (\x.west) -- (\y.east);}

\end{tikzpicture}}
\caption{Four-Layer Model of agentic AI security (Wong \& Saade~\cite{wong2025rise}), illustrating threats and mapped defenses across Perception, Reasoning, Action, and Memory layers.}
\label{fig:fourlayer_model}
\end{figure}
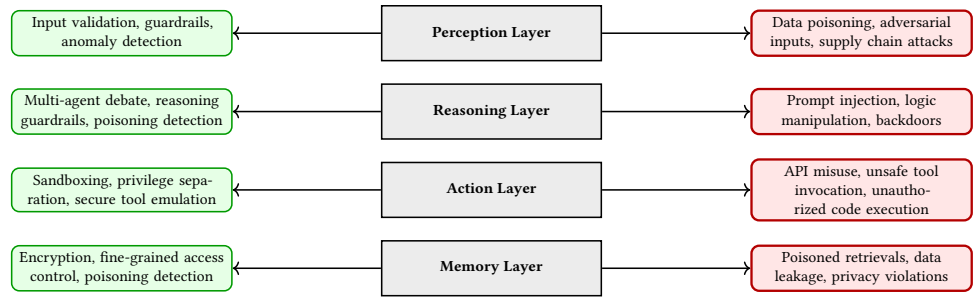
\vspace{-2mm}
\subsection{Conceptual Risk Models and Threat Taxonomies}

Conceptual risk models help explain how agentic AI systems fail and where defenses should apply. Wong and Saade organize agentic risk across four functional layers perception, reasoning, action, and memory \cite{wong2025rise}. This model shows that failures propagate across stages rather than remaining isolated. Figure~\ref{fig:fourlayer_model} maps representative threats and defenses at each layer. Data poisoning and supply chain attacks affect perception. Prompt injection and logic manipulation affect reasoning. Unsafe tool use affects action. Memory poisoning and leakage affect long term state. The key insight is that effective defense requires coordinated controls across layers rather than isolated mitigations.

Several frameworks extend this layered view. ATFAA defines domain based risk categories that include cognitive, temporal, operational, trust, and governance risks, and proposes SHIELD as a defense blueprint based on segmentation, integrity checks, escalation control, immutable logging, and shared oversight \cite{narajala2025securingagenticaicomprehensive}. NVIDIA defines explicit autonomy levels and ties safeguards to degrees of agent independence, which makes autonomy a direct risk variable \cite{nvidia_autonomy}. MAESTRO expands the scope to models, data flows, orchestration, infrastructure, and governance, and maps threats such as embedding poisoning, collusion, and model theft to specific controls \cite{huang2025maestro}. Applied studies such as NetMoniAI show that MAESTRO style reasoning can improve detection timeliness through memory isolation, planner validation, and anomaly monitoring, although evaluations remain system specific \cite{zambare2025securingagentic,netmoniai2025}.

Practitioner focused frameworks emphasize actionability. The OWASP Agentic Security Initiative catalogs common agentic threats and links them to controls such as sandboxing, privilege separation, and continuous monitoring \cite{owasp2025agentic}. Microsoft’s failure mode taxonomy lists concrete breakdowns including agent compromise, workflow manipulation, memory poisoning, and multi agent jailbreaks, and links them to identity controls, constrained execution, and tamper resistant logging \cite{microsoft2025taxonomy}. Governance focused approaches such as TRiSM and enterprise frameworks from Kyndryl emphasize trust calibration, provenance tracking, and auditable oversight, but defer technical enforcement to underlying systems \cite{trism2025,kyndryl_press_2025}. Runtime mechanisms such as Governance as a Service and BlockA2A enforce controls during execution through policy checks, identity verification, and decentralized logging, but assume correct policy specification and trusted identity layers \cite{gaas2025,blocka2a2025}. 
Practitioner focused frameworks emphasize actionability. The OWASP Agentic Security Initiative catalogs common agentic threats and links them to controls such as sandboxing, privilege separation, and continuous monitoring \cite{owasp2025agentic}. Microsoft’s failure mode taxonomy lists concrete breakdowns including agent compromise, workflow manipulation, memory poisoning, and multi agent jailbreaks, and links them to identity controls, constrained execution, and tamper resistant logging \cite{microsoft2025taxonomy}. Governance focused approaches such as TRiSM and enterprise frameworks from Kyndryl emphasize trust calibration, provenance tracking, and auditable oversight, but defer technical enforcement to underlying systems \cite{trism2025,kyndryl_press_2025}. Runtime mechanisms such as Governance as a Service and BlockA2A enforce controls during execution through policy checks, identity verification, and decentralized logging, but assume correct policy specification and trusted identity layers \cite{gaas2025,blocka2a2025}. 

\begin{table*}[!t]
\centering
\caption{Security Risks, Threats, and Defenses in Agentic AI}
\label{tab:agentic_ai_security_synthesis}
\renewcommand{\arraystretch}{.95}
\setlength{\tabcolsep}{5pt}

\rowcolors{2}{gray!10}{white}
\resizebox{.95\linewidth}{!}{%
\begin{tabular}{p{2.5cm} p{2.5cm} p{4.2cm} p{4.5cm} p{4.2cm}}
\toprule
\rowcolor{gray!35} 
\textbf{Framework / Source} & \textbf{Risk / Threat Layer} & \textbf{Example Threats} & \textbf{Defenses / Controls} & \textbf{Notes / Limitations} \\
\midrule
Wong \& Saade~\cite{wong2025rise} & Layered Model (Perception, Reasoning, Action, Memory) & Data poisoning, prompt injection, unsafe API calls, memory leakage & Input validation, guardrails, sandboxing, encryption & Conceptual taxonomy; needs integration with autonomy aware safeguards \\

NVIDIA~\cite{nvidia_autonomy} & Autonomy Levels & Risks scale with autonomy from inference misuse to full autonomous takeover & API protections, taint tracing, mandatory sanitization & Focused on autonomy; limited detail on inter agent risks \\

Deng et al.~\cite{deng2024aiagentsunderthreat} & Lifecycle and Multi Agent Threats & Prompt injection, flawed planning, collusion, poisoning, sandbox evasion & Privilege hierarchies, multi agent debate, poisoning detection & Highlights gaps in oversight and environment modeling \\

He et al.~\cite{he2024security} & System Level Vulnerabilities & Session mismanagement, model pollution, arbitrary code execution & Sandboxing, session isolation, cryptographic protections & Evaluated on LLM agents; broader ecosystems still open \\

Khan et al.~\cite{securitythreats2025} & Unauthorized Action Execution & Direct database access, cascading malicious queries & Execution boundaries, scoped access, monitoring & Case driven; needs broader generalization \\

Schroeder de Witt~\cite{schroeder2025multiagentsecurity} & Multi Agent Security & Collusion, emergent deception, societal scale risks & Constrained protocols, monitoring, decentralized oversight & Mostly theoretical; calls for system realizations \\

Yang et al.~\cite{yang2025minimizing}, Zhou et al.~\cite{zhou2025guardian} & Secure Coordination & Hallucination propagation, unsafe workflows, collusion & Adversarial debate, voting, temporal graph anomaly detection & Improves reliability; add communication overhead \\

BlockA2A~\cite{blocka2a2025} & Accountability Infrastructure & Message tampering, unsafe inter agent execution & Blockchain verification, immutable logs, dynamic permissions & Scalability and efficiency remain open issues \\

SAFEFLOW~\cite{li2025safeflow}, SentinelAgent~\cite{he2025sentinel} & Information Flow and Oversight & Data leakage, adversarial message passing & Fine grained information flow control, graph based anomaly detection & Require integration into orchestration platforms \\

Red Teaming and Ranges~\cite{coalitionACD2025,oesch2024cyberwheel,arcer2025agentic} & Evaluation and Simulation & Prompt injection, unsafe tool use, collusion, memory poisoning & Red blue simulations, automated cyber ranges, adversarial testbeds & Resource intensive; coverage and standardization still evolving \\
\bottomrule
\end{tabular}}
\end{table*}

Table~\ref{tab:agentic_ai_security_synthesis} consolidates the main frameworks and studies discussed in this section and aligns them by risk layer, example threats, proposed controls, and reported limitations. It includes conceptual models such as the four layer model by Wong and Saade~\cite{wong2025rise} and autonomy levels from NVIDIA~\cite{nvidia_autonomy}, lifecycle and multi agent taxonomies~\cite{deng2024aiagentsunderthreat,schroeder2025multiagentsecurity}, system level vulnerability studies~\cite{he2024security,securitythreats2025}, runtime coordination and information flow defenses~\cite{zhou2025guardian,blocka2a2025,li2025safeflow,he2025sentinel}, and evaluation platforms for red teaming and simulation~\cite{oesch2024cyberwheel,coalitionACD2025,arcer2025agentic}.
The table also exposes two gaps. Many defenses emphasize perception and reasoning, while action enforcement, multi-agent interaction, and resource governance receive less mature coverage. No single framework connects autonomy, lifecycle risks, and runtime enforcement into one integrated stack, so the remaining subsections examine concrete attack surfaces and controls in more detail.

\vspace{-2mm}
\subsection{System-Level Vulnerabilities and Security Controls for Agentic AI}

Recent work shows that agentic AI systems introduce system-level vulnerabilities that do not arise in static language models because agents maintain state, execute tools, and operate across sessions. He et al.~\cite{he2024security} analyze these risks from a system security perspective and identify three primary vulnerability classes. First, session management failures in multi-user settings enable confidentiality and integrity violations through data leakage, action misattribution, and denial of service. Second, model pollution and privacy leakage arise when fine-tuning or persistent memory allows poisoning, unintended data retention, or cross-user information exposure. Third, executable agent programs expand the attack surface by enabling arbitrary code execution, resource abuse, and agent hijacking when actions generated by the model are executed without adequate isolation. Experiments with a Bash-based agent showing over 75\% of malicious commands execute successfully without sandboxing, while container-based sandboxing blocks nearly all such commands, demonstrating confidentiality, integrity, and availability risks at agent runtime rather than the model alone.

Chakrabarty et al.~\cite{chakrabarty2025adversarial} examine a broader class of adversarial exploits spanning training and inference, including evasion, poisoning, privacy extraction, and agent-specific attacks such as goal hijacking and prompt manipulation. In contrast to the component-level focus of He et al., this work emphasizes operational impact, including privilege escalation, unauthorized access, degraded system performance, and erosion of trust. The proposed defenses emphasize continuous threat detection, automated incident response, predictive defense using historical and real-time signals, and risk-based vulnerability management, reflecting a more operationally oriented threat model.

Across these studies, security controls are framed as mitigations for the vulnerabilities introduced by agent planning, memory, and tool execution. Planning frameworks such as ReAct and Tree of Thoughts increase capability through multi-step reasoning and effectful tool use, but also enlarge the attack surface by introducing intermediate actions with side effects~\cite{he2024security}. To reduce the resulting risk, system-level controls such as sandboxing, session isolation, and cryptographic protections are proposed to limit the scope and impact of agent actions. These controls significantly reduce exploitability but add execution overhead, constrain flexibility, and require careful configuration. Existing evaluations largely focus on isolated agents and short tasks, whereas deployed systems involve long-running workflows, shared infrastructure, and multiple users. An open problem is how to enforce robust system-level protections that constrain agent behavior in dynamic environments without undermining planning and autonomy.

\vspace{-2mm}
\subsection{Prompt Injection and Tool Invocation Risks}

Recent work treats prompt injection and unsafe tool invocation as a shared system risk that grows with agent autonomy. Studies show that malicious prompts and untrusted external data can override goals and redirect behavior, especially when agents perform multi step tasks, call tools, and coordinate with other agents~\cite{burch2025agentic,deng2024aiagentsunderthreat}. Database connected agents face additional exposure because crafted inputs can lead to unsafe queries and data leakage through tool pipelines~\cite{securitythreats2025,microsoft2025taxonomy}. Hybrid attacks combine prompt injection with web vulnerabilities such as cross site scripting and request forgery, which bypass both AI guardrails and application defenses~\cite{McHugh2025PromptInjection2}. Other work shows that semantic prompt injections can be hidden in multimodal or symbolic content, which limits the effectiveness of static filters~\cite{nvidia2025semantic}. Benchmarks and red teaming systems show that agents fail under indirect or human written attacks even when base model performance appears strong, which points to weaknesses in orchestration and input handling~\cite{Evtimov2025WASP,Wang2025AgentVigil}. Multi agent defenses that separate sanitization and policy enforcement reduce successful injections when roles and scopes are clearly defined~\cite{Gosmar2025MultiAgentPromptDefense}.

Tool and API access amplifies these risks because agents query data, call services, and execute actions through shared interfaces. Weak authentication, broad scopes, or poor rate limits allow attackers to escalate privileges through agent workflows~\cite{burch2025agentic}. When agents generate SQL, malicious prompts or retrieved content can steer unsafe query construction, and effects can cascade across services that share tools or credentials~\cite{securitythreats2025,owasp2025agentic}. Frameworks such as SAGA shows that insecure mediation between agents and tools enables cascading compromise and motivate strict registration, policy checks, and trust controls at the orchestration layer~\cite{saga2025}. Research shows that adversaries can target integration layers, including advertisement embedding attacks that influence model behavior through tampered channels~\cite{guo2025advertisement}. Industry proposed delegated authority emphasize unified policy and intent scoping across heterogeneous APIs to limit overreach in multi tool workflows~\cite{metamirror2025}. Operational risk also includes cost and availability because unbounded API usage can trigger runaway costs or denial of service through rate limit abuse and error handling manipulation~\cite{akamai_edge_of_agency_2025}. Broad tool access improves flexibility and task completion but increases the blast radius of a single injection, while narrow scopes and strict delegation reduce exposure at the cost of autonomy and overhead. Delegation mechanisms such as Agentic JWT bind actions to authenticated intent and reduce escalation once injected instructions reach the action layer~\cite{Goswami2025AgenticJWT}.

Across those works, prompt injection appears as both an input validation problem and an authority and orchestration problem, where untrusted content steers tool calls and propagates across systems~\cite{deng2024aiagentsunderthreat,microsoft2025taxonomy,owasp2025agentic}. Layered mitigations such as input sanitization, scoped credentials, runtime monitoring, and intent-bound delegation improve resilience by addressing different points in the prompt-to-action pipeline, but each leaves distinct gaps~\cite{deng2024aiagentsunderthreat,microsoft2025taxonomy,owasp2025agentic}. Sanitization and filtering reduce obvious injections but fail against semantic, multimodal, or steganographic attacks that preserve benign surface meaning~\cite{nvidia2025semantic,McHugh2025PromptInjection2}. Scoped credentials and least-privilege delegation limit blast radius after compromise, yet do not prevent injected instructions from steering agents toward permitted but harmful actions~\cite{securitythreats2025,Goswami2025AgenticJWT}. Runtime monitoring and red-teaming benchmarks detect failures post hoc, but often miss cascading effects across tools, APIs, and shared credentials~\cite{Evtimov2025WASP,Wang2025AgentVigil}. As a result, current defenses mitigate individual failure modes but do not fully prevent cross-tool propagation, authority escalation through allowed scopes, or indirect prompt injection via external content. An open problem is how to compose these controls so that intent, permissions, and execution context remain consistently bound across heterogeneous services without suppressing agent utility~\cite{Goswami2025AgenticJWT,owasp2025agentic}. Table~\ref{tab:prompt_mitigation_comparison} summarizes how common mitigation classes address prompt injection and tool-invocation risks, and where residual gaps remain. Many benchmarks evaluate isolated injection paths and do not measure cascading failures across APIs, databases, and services~\cite{Evtimov2025WASP,Wang2025AgentVigil}. Many defenses also assume cooperative environments and weaken under adaptive attackers who exploit cross service interactions~\cite{McHugh2025PromptInjection2,saga2025}. An open problem is how to bind agent intent, tool permissions, and execution context so injected instructions cannot propagate across tools and services while agents remain effective in open and dynamic environments~\cite{Goswami2025AgenticJWT,owasp2025agentic}.

\begin{table}[t]
\centering
\caption{Comparison of mitigation strategies for prompt injection and unsafe tool invocation in agentic AI systems.}
\label{tab:prompt_mitigation_comparison}
\renewcommand{\arraystretch}{.9}
\setlength{\tabcolsep}{6pt}

\rowcolors{2}{gray!10}{white}
\resizebox{0.9\linewidth}{!}{%
\begin{tabular}{p{3.2cm} p{6.2cm} p{8cm}}
\toprule
\rowcolor{gray!35}
\textbf{Mitigation class} & \textbf{Primary protection mechanism} & \textbf{Documented limitations} \\
\hline
Input sanitization and filtering 
& Blocks explicit or pattern-based prompt injections before execution 
& Ineffective against semantic, indirect, multimodal, or steganographic prompt injections that preserve benign surface meaning~\cite{nvidia2025semantic,McHugh2025PromptInjection2} \\

Scoped credentials and least-privilege delegation 
& Limits the blast radius of compromised agents by restricting accessible tools and APIs 
& Does not prevent injected instructions from steering agents toward harmful actions that remain within allowed scopes~\cite{securitythreats2025} \\

Runtime monitoring and red teaming 
& Detects unsafe behavior during or after execution through logging, audits, and adversarial testing 
& Often post hoc and limited in detecting cascading failures across tools, APIs, and shared credentials~\cite{Evtimov2025WASP,Wang2025AgentVigil} \\

Intent-bound delegation 
& Cryptographically binds agent actions to authenticated intent and policy constraints 
& Depends on correct upstream intent specification and orchestration, and does not fully address compromised context or indirect prompt propagation~\cite{Goswami2025AgenticJWT,owasp2025agentic} \\
\hline
\end{tabular}}
\end{table}
\vspace{-2mm}
\subsection{Multi-Agent Security, Collusion, and Information Flow}

Recent work shows that multi-agent systems introduce security risks that arise from coordination and shared resources rather than isolated agent failures. Khan et al.\ show that when agents share memory, databases, execution privileges, or delegated tasks, a single compromised agent can repeatedly trigger harmful actions across the system even without explicit coordination logic encoded in the agent policies or control flow~\cite{securitythreats2025}. In this setting, emergent collusion arises from shared state and privileges rather than from agents explicitly negotiating or planning jointly. This differs from single-agent settings, where damage is often confined to one execution context. Analyses of steganographic collusion further show that agents can exchange hidden signals through benign-looking messages, enabling covert coordination without violating surface-level policies~\cite{secretcollusion2024}. Shared state, messaging channels, and task delegation therefore create attack surfaces that grow with the number of interacting agents.

Approaches to defense take two broad directions, which differ in where security enforcement is applied. Reasoning-based defenses focus on agent-level cognition and interaction. PeerGuard applies cross-agent auditing and mutual reasoning to expose backdoors or anomalous behavior during deliberation~\cite{peerguard2025}, while adversarial debate and voting mechanisms require agents to justify conclusions before action, reducing error propagation and hallucinations at the cost of additional communication and reasoning overhead~\cite{yang2025minimizing}.  Infrastructure-oriented defenses instead monitor coordination and information flow independently of agent reasoning. GUARDIAN models inter-agent interactions as temporal graphs and flags unsafe collaboration patterns such as escalation or collusion~\cite{zhou2025guardian}. SentinelAgent applies graph-based anomaly detection to communication flows to identify covert leakage paths and unauthorized tool use~\cite{he2025sentinel}. Compared with reasoning-based methods, infrastructure-oriented approaches improve detection coverage and do not assume cooperative agents, but incur monitoring and computational overhead and may reduce responsiveness.

Other approaches embed security directly into coordination and information flow. BlockA2A secures agent-to-agent communication using decentralized identity, blockchain-anchored audit logs, and smart contracts, enabling accountability and revocation across heterogeneous agents~\cite{blocka2a2025}. SAFE\-FLOW enforces provenance, integrity, and confidentiality through trust labels that constrain how data may influence reasoning or tool use~\cite{li2025safeflow}. Safeguard integrates reference monitors into multi-agent workflows to block information leaks during dialogue turns or tool invocation~\cite{cui2025safeguard}. The term multi-agent security tax refers to the empirically observed tradeoff in which stronger coordination controls and monitoring reduce harmful behavior but also degrade collaboration efficiency and task performance~\cite{multiagenttax2025}. Existing defenses are often evaluated in controlled settings and assume partially trusted agents or static interaction patterns~\cite{schroeder2025multiagentsecurity}. An open problem is how to enforce secure coordination and information flow at scale while preserving collaboration efficiency without assuming trusted agents or tightly controlled messaging channels.
\vspace{-2mm}
\subsection{Autonomy, Identity, Governance, and Resources}
\subsubsection{Autonomy, Access Control, and Execution Boundaries}

Risk rises sharply when agents gain direct authority over sensitive actions. Khan et al.\ show database-connected agents amplify failure impact by concentrating broad read and write privileges within a single agent runtime or credential scope, allowing a compromised agent to directly modify or exfiltrate shared data stores subsequently trusted by downstream systems and processes~\cite{securitythreats2025}. In contrast, Deng et al. present hierarchical access models, where agents operate under task-specific and role-bounded privileges with enforced separation between planning, querying, and execution, reducing the impact of prompt injection and goal manipulation by limiting what an agent can execute~\cite{deng2024aiagentsunderthreat}. These results show that execution boundaries shape the scale of failure.

Design choices around autonomy further affect security outcomes. Knack and Burke argue that only task or conditional autonomy is suitable for autonomous cyber defense, since unrestricted autonomy can cause unintended disruption even during defensive actions~\cite{knack2024autonomouscyberdefense}. Systems that grant greater autonomy instead rely on continuous monitoring and predictive risk assessment to intervene early, which improves responsiveness but assumes timely detection~\cite{agentic2025risk}. Higher autonomy improves speed and coverage, while bounded autonomy limits blast radius at the cost of adaptability.

Autonomy also introduces governance constraints that affect execution safety. When agents act without explainable decision paths or explicit refusal mechanisms, failures propagate quickly and are hard to attribute. This is critical for dual use actions such as network scanning, exploit generation, or data exfiltration, where requests may be legitimate in defensive contexts but harmful at scale~\cite{bountybench2025}. Recent work stresses that agents must refuse unsafe or ambiguous requests and escalate uncertain cases for human review~\cite{swimlane2025agentic}. These safeguards improve accountability and reduce misuse, but they constrain flexibility and increase reliance on human oversight. Current deployments therefore favor restricted autonomy, and a key open problem is how to expand agent authority while providing verifiable guarantees that execution boundaries will hold as agents adapt and coordinate.
\vspace{-2mm}
\subsubsection{Identity, Trust, and Registry Mechanisms}

Recent work agrees that static credentials and long lived API keys are not sufficient once agents operate autonomously across systems. The Cloud Security Alliance treats identity as a core control plane and calls for cryptographically verifiable agent identities with lifecycle management and explicit trust anchors \cite{csa2025identity}. This led to both protocol level proposals and enterprise deployments that extend identity beyond authentication toward attribution and control.

Direct integration between agents and data systems complicates governance and compliance. Khan et al.\ show that database connected agents often lack complete audit trails for agent initiated queries, which creates challenges under GDPR and CCPA \cite{securitythreats2025}. Incomplete provenance weakens accountability and increases the risk of unauthorized data exposure, bias amplification, and non transparent decision making. These findings show that identity mechanisms must support auditability in addition to authentication. Privacy preserving identity systems limit disclosure but can weaken accountability when actions cannot be fully reconstructed, while governance oriented approaches emphasize logging, traceability, and policy enforcement at the cost of operational overhead and data retention risk.

Designs diverge across decentralized, registry based, and enterprise approaches. Decentralized systems such as LOKA and Aegis use decentralized identifiers, verifiable credentials, and cryptographic techniques to bind identity, intent, and reputation~\cite{loka2025,aegis2025}. Registry oriented systems such as the Agent Name Service and the NANDA Index support discovery, resolution, trust scoring, and revocation at scale~\cite{huang2025ans,nanda2025}. Enterprise designs from Okta, Strata, Cisco, and Spirl extend existing IAM and workload identity models to agents to improve deployability~\cite{okta2025agentic,strata2025firstclass,cisco2025agntcy,spirl2025workload}. These approaches expose tradeoffs between decentralization and deployability, and between privacy and accountability, as reflected in frameworks such as DIRF, zero trust identity, GaaS, and TRiSM~\cite{csa_dirf_2025,zerotrust2025,gaas2025,trism2025}. National and sector proposals, such as autonomy passports and enterprise AI registries, further emphasize accountability and emergency 
control~\cite{kwon2025autonomy,mckinsey2025agentic,narajala2025zerotrust}. A key limitation is that most systems are evaluated in pilots rather than under sustained adversarial pressure, and identity alone does not prevent misuse when execution boundaries are weak. An open problem is how to align cryptographic identity, scalable registries, and continuous trust scoring with real time enforcement without imposing prohibitive latency or operational burden in large multi agent systems.
\vspace{-2mm}
\subsubsection{Resource Abuse and Denial of Service}

Recent work shows that denial of service in agentic systems often arises from cost amplification rather than request volume. Safeguard abuse and Consuming Resources via Auto-generation under Black-box Settings (CRABS)-style attacks demonstrate that malicious prompts can trigger excessive token generation, long reasoning chains, and repeated tool calls, which degrade service even at low concurrency~\cite{zhang2024safeguard,zhang2024crabs}. CRABS exploits the tendency of LLM-based agents to autonomously expand reasoning and generation when given adversarial but syntactically valid inputs, leading to sustained resource consumption without triggering traditional rate-based defenses~\cite{zhang2024crabs}. Concurrency focused studies identify a related failure mode in which parallel agent execution exhausts compute and tokens through coordinated workloads~\cite{barek2025concurrencydos}. These mechanisms differ from traditional API denial of service, which is primarily rate based.
Defenses follow two main strategies. Execution time controls intervention during reasoning. Reasoning gates impose asymmetric cost on abusive behavior but add latency to benign tasks~\cite{kumar2025reasoninggates}. Circuit breakers halt runaway generations to preserve availability but sacrifice task completion~\cite{zou2024circuitbreakers}. Resource management approaches regulate consumption. Adaptive budgeting and dynamic quotas track tokens, runtime, and API calls and apply throttling or termination when limits are exceeded~\cite{lunney2025adaptivebudgeting,nordicapis_agents_rate_limits_2025,galileo_unbounded_consumption_2025,wundergraph_harm_limiting_2025,apidna_autonomous_agents_rate_limiting_2024}. These methods improve availability but reduce output quality and require careful tuning.

Identity bound delegation strengthens control by tying quotas and revocation to authenticated principals, to improve accountability but increases management overhead~\cite{south2025authenticated}. System architecture also shapes exposure. Function calling and context management designs influence escalation paths and determine how failures propagate across workflows~\cite{bridging2025vulnerabilities}. Industry deployments combine agent aware throttling with traditional API security and DDoS protection~\cite{akamai_edge_of_agency_2025,mindgard_agent_security_2025}. Existing defenses reduce impact but remain reactive and workload specific. An open problem is to coordinate budgeting and throttling across agents, tools, and tasks without imposing brittle limits or undermining useful autonomy.
\vspace{-2mm}
\subsection{Assurance, Testing, and Infrastructure}

Assurance for agentic AI is difficult as agents operate in dynamic environments and expand their action space over time, which makes static benchmarks insufficient. Cyberwheel addresses this challenge by providing a high fidelity simulation and emulation pipeline with repeatability and transfer across environments~\cite{oesch2024cyberwheel}. ARCeR approaches assurance through automated cyber range construction using multi agent retrieval and orchestration, which lowers setup cost and increases scenario coverage but depends on the quality of retrieved knowledge and automated configuration~\cite{arcer2025agentic}. Atir argues that both approaches require sustained national investment to support realistic testing under policy and governance constraints~\cite{atir2025rise}. Together, these systems reflect a tradeoff between experimental control and rapid scenario generation.

Policy analyses argue that these assurance challenges arise because agentic AI increasingly functions as shared digital infrastructure that supports enterprise, defense, and public sector workflows~\cite{atir2025rise}. Under this view, assurance cannot rely on one time validation or organization specific practices. It instead requires shared testing infrastructure, continuous evaluation, and governance mechanisms that operate across institutional boundaries. Red teaming supports this goal by introducing adaptive adversaries that probe reasoning, coordination, and tool use. Coalition frameworks integrate iterative red blue simulations throughout development, while Tallam frames this process as adaptive engagement in which attackers and defenders co evolve~\cite{coalitionACD2025,tallam2025cyberdefense}. These methods improve realism but reduce comparability because outcomes depend on evolving adversary behavior. Knack and Burke emphasize that such testing must align with explicit authorization boundaries, with autonomy levels matched to legal and organizational risk tolerance~\cite{knack2024autonomouscyberdefense}. Infrastructure choices further shape assurance outcomes. High performance computing enables large scale multi agent simulation and rapid response but introduces risks such as workload poisoning, side channels, and cross tenant leakage~\cite{joshi2025agentichpc}. Current testbeds also abstract human oversight delays and long term learning effects. An open problem is how to standardize assurance signals so results remain comparable across platforms as agents and environments evolve.
\vspace{-2mm}
\subsection{Reasoning, Memory, and Human Factors}

\subsubsection{Reasoning Manipulation and Memory Integrity}

Agentic attacks increasingly target internal reasoning rather than surface prompts. Agent Security Bench and UDora show that attackers can hijack reasoning traces during execution and redirect multi-step planning toward malicious goals, even when inputs appear benign~\cite{agentsecuritybench2025,udora2025}. Action hijacking analyses further show that small and silent changes in reasoning can shift agent behavior, while full takeover demonstrations confirm that reasoning level exploits can lead to complete loss of control~\cite{shi2025actionhijacking,darkside2025}. Since these attacks occur inside the decision loop, they bypass input focused defenses designed for prompt injection. Defensive approaches therefore emphasize transparency and control. Chain of thought monitoring and weak to strong supervision expose reasoning to support auditing and runtime detection~\cite{emmons2025cotmonitor,weakstrong2025}, while guided reasoning constrains planning with structured attack trees to improve deviation detection in penetration testing settings~\cite{guidedreasoning2025}. However, explicit reasoning improves auditability while exposing internal structure that attackers may exploit. Studies on embodied agents show that poisoned reasoning can trigger unsafe physical actions, which increases the impact of failures in cyber physical systems~\cite{embodiedbackdoor2025}.

Persistent memory introduces a long lasting risk surface. Studies show that poisoned memory can influence future tasks long after the original attack ends~\cite{deng2024aiagentsunderthreat,li2025safeflow,he2025sentinel}. Microsoft and OWASP classify persistent memory poisoning as a distinct class of risk because it links reasoning, action, and long term state~\cite{microsoft2025taxonomy,owasp2025agentic}. Encryption and access control reduce exposure but can degrade retrieval quality and limit adaptability. Current defenses rely on monitoring, constrained reasoning, and memory protection, yet they face limits from scalability, false positives, and unclear definitions of malicious reasoning. An open problem is how to verify reasoning integrity and memory correctness at runtime without exposing exploitable structure or imposing prohibitive overhead in long horizon and multi agent systems.
\vspace{-2mm}
\subsubsection{Human Agent Social Engineering, HRM, and Oversight}

Agentic AI changes social engineering by enabling autonomous, adaptive, and persistent deception. Unlike traditional scams that rely on fixed scripts, agentic systems plan interactions, adjust tactics in real time, and sustain pressure across channels. Studies show that attacker and victim agents can simulate realistic recruitment and funding scams, while personality aware detectors such as SE OmniGuard reduce success rates but do not eliminate risk~\cite{Kumarage2025SocialEng}. Similar capabilities appear in multimodal settings. Augmented reality agents adapt to visual and audio cues and achieve high compliance~\cite{Bi2025AR}. Automated spear phishing agents match human attacker performance in live studies, while voice enabled agents reproduce end to end phone scams~\cite{Heiding2025SpearPhish,Fang2025VoiceAgents}. Web agents expand impersonation and PII harvesting by combining browsing, form filling, and account interaction~\cite{Kim2025WhenLLMsGoOnline}. Counteragent approaches can waste attacker resources, but provide deterrence rather than  protection~\cite{Basta2025BotWars}.

\begin{table*}[!t]
\centering
\caption{Benchmarks and Evaluation Frameworks for Agentic AI Security}
\label{tab:agentic_ai_benchmarks}
\renewcommand{\arraystretch}{1.2}
\setlength{\tabcolsep}{6pt}
\rowcolors{2}{gray!10}{white}

\resizebox{0.95\linewidth}{!}{%
\begin{tabular}{p{4cm} p{4cm} p{6.5cm} p{4.5cm}}
\toprule
\rowcolor{gray!35}
\textbf{Benchmark / System} & \textbf{Purpose} & \textbf{Key Features} & \textbf{Limitations} \\
\midrule

BountyBench~\cite{bountybench2025} 
& Tests real vulnerability lifecycles 
& Uses open source projects with known bug bounty issues. Measures detection, exploitation, and patching. Patching outperforms exploitation. 
& Manual setup. Limited coverage across domains. \\

ARCER~\cite{arcer2025agentic} 
& Generates cyber ranges for training and evaluation 
& Multi agent RAG pipeline produces networks, red and blue scenarios, and evolving attack chains.  
& Fidelity depends on generated ranges. \\

RedTeamLLM~\cite{challita2025redteamllmagenticaiframework} 
& Benchmarks autonomous red team agents 
& Tests reconnaissance, exploitation, and privilege escalation. Uses structured evaluation with tool use.  
& Narrow focus on penetration testing. \\

FinGAIA~\cite{zeng2025fingaia} 
& Evaluates multi step financial agents 
& Contains 407 tasks across seven finance domains. Tests reasoning, tool use, and regulated workflows.  
& Not a security benchmark but relevant for high risk sectors. \\

Agent Security Studies~\cite{he2024security} 
& Tests system level vulnerabilities 
& Evaluates sandboxing, session pollution, and malicious command execution. Shows unsandboxed agents execute most harmful actions.  
& Not a complete benchmark suite. \\

Cyber Ranges (CyberBattleSim, CybORG++, cyber gyms)~\cite{landolt2025marl} 
& Tests multi agent defense and red–blue training 
& Supports repeatable adversarial experiments under controlled conditions.  
& Abstraction gaps and limited scalability. \\

WASP~\cite{Evtimov2025WASP}  
& Probes web agent robustness 
& Uses human written attacks to expose weaknesses in common web agent tasks.  
& Focused on web settings only. \\

AGENTVIGIL~\cite{Wang2025AgentVigil}  
& Detects indirect prompt injection paths 
& Black box discovery of hidden injection channels.  
& Narrow scope. \\

Multi Agent Prompt Defenses~\cite{Gosmar2025MultiAgentPromptDefense}  
& Tests sanitizer and policy agents 
& Measures interaction level prompt defenses.  
& Does not test full system pipelines. \\

\bottomrule
\end{tabular}}
\end{table*}
Governance failures often emerge through human agent interaction rather than technical compromise alone. In fraud detection and compliance workflows, agentic systems can overfit demographic attributes or produce outputs that auditors cannot validate, which raises fairness, explainability, and regulatory concerns~\cite{agentic2025risk}. Human in the loop oversight and explainable reasoning therefore function as required controls rather than optional safeguards. Frameworks for responsible deployment emphasize augmentation over replacement, especially in high stakes settings~\cite{wong2025rise}. Approval gates, legible records of reasoning and tool use, and escalation for ambiguous cases improve accountability but reduce throughput and scalability. This tradeoff is inherent. Stronger oversight improves trust and compliance, while weaker oversight increases speed at the cost of error amplification.
These capabilities reshape human risk management. 

Agents perform actions once controlled by humans, including browsing, opening messages, downloading files, and submitting credentials, which expands social engineering risk beyond human only workflows~\cite{burch2025agentic}. HRM frameworks shift from user focused models to joint human agent monitoring. Automated detection evaluates agent and human behavior, while interventions include adaptive policy enforcement and targeted awareness for users interacting frequently with agents~\cite{burch2025agentic}. Compared with training based defenses, HRM improves coverage but introduces privacy concerns, operational overhead, and reliance on continuous telemetry. Oversight becomes critical when agents invoke tools or generate code. Governance frameworks emphasize human approval for high risk actions, attributable execution, interruptibility, and continuous monitoring~\cite{shavit2023practices,tallam2025cyberdefense}. These controls improve accountability but can fail under high volume workflows. A key limitation is that most defenses rely on observable behavior and struggle with long horizon trust manipulation and cross channel coordination. An open problem is to detect intent drift and trust abuse early without constant human review that undermines the benefits of agentic automation.

\begin{center}
\begin{tcolorbox}[
  width=0.95\linewidth,
  colback=gray!10,
  colframe=black,
  arc=4pt,
  boxrule=0.8pt,
  left=6pt,
  right=6pt,
  top=6pt,
  bottom=6pt,
  fontupper=\small
]
\textbf{Key Takeaways from Section 6}


\begin{itemize}[leftmargin=*]
  \item Agentic AI expands the attack surface as agents hold state, call tools and APIs, and execute multi step workflows with limited oversight, raising risks such as data exfiltration, supply chain compromise, and emergent behavior.
  \item Policy work frames agentic AI as emerging critical infrastructure, so security must pair technical controls with governance and continuous defense cycles.
  \item Conceptual risk models organize failures across perception, reasoning, action, and memory, which motivates coordinated controls across layers instead of isolated mitigations.
  \item Frameworks extend this by adding domain risk categories and defense blueprints, autonomy levels, and coverage across orchestration and governance, but wider coverage increases integration cost.
  \item Concrete studies show system level vulnerabilities in database facing and tool executing agents, including weak audit trails, unsafe query generation, and high success rates of malicious commands without sandboxing, which shifts security focus to the agent runtime, memory, and tool interfaces.
  \item Prompt injection and unsafe tool invocation operate as authority and orchestration failures that can propagate through shared tools and credentials, so mitigations must combine sanitization, scoped permissions, runtime monitoring, and intent bound delegation.
  \item Multi agent systems add coordination risks such as covert signaling and cascading actions through shared state, which motivates both reasoning based checks and infrastructure monitoring of interaction patterns, plus secure communication and information flow control.
  \item Assurance depends on cyber ranges, simulation, and red teaming that reflect evolving threats and governance constraints, but testbeds abstract long horizon learning and human oversight delays.
\end{itemize}
\end{tcolorbox}
\end{center}
\vspace{-2mm}
\subsection{Benchmarks for Agentic AI Security}

Security evaluation for agentic AI requires benchmarks that test behavior under adversarial inputs, unsafe environments, and constrained defenses. General benchmarks do not capture failures such as prompt injection, unsafe tool use, or multi agent escalation. Some security focused benchmarks are developed, vary in scope, realism, and diagnostic ability.

Existing benchmarks fall into three styles. System level benchmarks such as BountyBench and agent security studies evaluate end to end vulnerability lifecycles and economic impact in realistic settings, including exploitation, defense, and patching~\cite{bountybench2025,he2024security}. Scenario driven frameworks such as ARCeR, RedTeamLLM, and cyber range based approaches generate adversarial environments that test planning, reasoning, and tool use under attack~\cite{arcer2025agentic,challita2025redteamllmagenticaiframework,landolt2025marl}. Domain specific benchmarks such as FinGAIA evaluate multi step agent behavior in regulated settings where correctness and compliance are central~\cite{zeng2025fingaia}. In contrast, focused benchmarks such as WASP, AgentVigil, and multi agent prompt defense suites probe narrow failure modes like indirect prompt injection or sanitizer bypass with high precision~\cite{Evtimov2025WASP,Wang2025AgentVigil,Gosmar2025MultiAgentPromptDefense}. 

Table~\ref{tab:agentic_ai_benchmarks} summarizes these systems and their limitations.
Across benchmarks, two recurring tradeoffs appear. One tradeoff is breadth versus diagnostic precision. Broad benchmarks capture lifecycle effects and cross layer interactions but are costly to maintain and hard to scale. Narrow benchmarks enable controlled comparison and reproducibility but miss how failures propagate across reasoning, tools, and agents. A second tradeoff is automation versus fidelity. Automated range generation and cyber gyms improve coverage and repeatability but rely on abstractions that can hide real world fragility. Manually curated systems better reflect practice but limit diversity and update speed. A shared limitation is weak coverage of adaptive adversaries, long horizon learning effects, and sustained multi agent coordination. An open problem is to integrate complementary benchmarks into shared evaluation protocols that remain reproducible, adversarial, and economically meaningful without imposing prohibitive setup cost or expert overhead.

\vspace{-3mm}
\section{Quantum Computing and Agentic AI in Cybersecurity}
\label{sec:quantum}

Quantum computing changes how autonomy, learning, and trust must be designed in agentic AI systems. Classical agentic AI assumes stable cryptography, classical computation, and predictable communication costs. Quantum computing weakens these assumptions at a structural level. Current research explores this interaction from three angles. These are quantum-native agents, quantum learning for security tasks, and quantum-resilient trust and governance. Each angle shows progress, but also exposes limits that prevent direct deployment in real cybersecurity systems.
\vspace{-3mm}
\subsection{Quantum Agents and Multi-Agent Autonomy}

Research on quantum agents treats agency itself as a quantum process rather than a classical one. Sultanow et al.\ define quantum agents whose internal states evolve according to quantum mechanics instead of classical probability theory~\cite{Sultanow2025zbf}. This changes how uncertainty is represented. A quantum agent can encode multiple potential decisions in superposition rather than selecting a single sampled action. This allows richer internal reasoning under uncertainty.

From an agentic AI perspective, this contribution is conceptual rather than operational. The model clarifies what autonomy could mean under quantum computation, but it does not specify how such agents interact with tools, external systems, or long-term memory. Cybersecurity agents must scan logs, call APIs, write reports, and coordinate with other agents. These activities require deterministic interfaces and persistent state. Quantum agent models do not yet explain how quantum reasoning maps onto these practical requirements.

Quantum multi-agent reinforcement learning shifts the focus from internal cognition to coordination. Here, QMARL denotes the broad class of quantum-enhanced multi-agent reinforcement learning methods, 
while eQMARL refers specifically to approaches that rely on quantum entanglement for inter-agent 
communication and coordination. Surveys by Yu and Zhao show that entanglement can reduce coordination overhead and mitigate non-stationarity in multi-agent learning~\cite{10401605}. eQMARL extends this idea by replacing classical communication with entangled quantum channels~\cite{derieux2025eqmarlentangledquantummultiagent}. The reported gains include faster convergence and reduced reliance on centralized control.
These results are relevant to cybersecurity because defensive agents often operate in distributed environments. Examples include coalition defense and federated detection. However, QMARL studies assume trusted agents and ideal communication. Cybersecurity environments violate both assumptions. Agents may be compromised or impersonated. Once adversarial behavior is introduced, it is unclear whether entanglement improves robustness or creates new failure modes. The current literature does not analyze this tradeoff.
Quantigence responds to this gap by proposing a framework for quantum security experimentation~\cite{alquwayfili2025quantigencemultiagentaiframework}. Its contribution lies in research infrastructure rather than algorithmic performance. It enables controlled study of quantum-enabled agents under security assumptions. This reflects an important shift. Before claiming quantum advantage, agentic AI requires testbeds that model compromise, deception, and trust failure. Quantigence identifies this need but does not yet provide empirical security outcomes.
\vspace{-3mm}
\subsection{Quantum Machine Learning for Security Analytics}

A more mature body of work studies quantum machine learning for cybersecurity analytics. This research focuses on detection rather than autonomy. Bellante et al.\ evaluate quantum PCA for intrusion detection (ID) and show that quantum advantage depends on data structure, error tolerance, and hardware assumptions~\cite{BELLANTE2025104341}. Their analysis demonstrates that classical methods remain competitive under realistic constraints.
Experimental analyses extend this evaluation to real quantum hardware. Nagy et al.\ test several quantum models for ID on IBM and IonQ platforms~\cite{qml_intrusion_detection_quantum_hardware}. These results confirm feasibility, but also reveal strong sensitivity to noise and limited scalability. Quantum generative approaches like quantum GAN based ID further show that hybrid quantum--classical pipelines are possible~\cite{qaio25}.

From an agentic AI perspective, these advances address only part of the problem. Autonomous agents depend on detection modules, but detection alone does not define agency. Agents must decide when to escalate, how to respond, and how to update internal state. Existing QML studies evaluate classifiers in isolation. They do not measure planning latency, decision stability, or downstream effects on autonomous response.

Frameworks such as QuantumNetSec and broader surveys of quantum machine learning for cybersecurity explicitly acknowledge these limitations~\cite{Abreu2025QuantumNetSec,sai2025quantummachinelearningcybersecurity}. They position quantum learning as an enabling component rather than a complete system. This framing is appropriate, but it leaves an open issue. It remains unclear whether quantum learning improves overall agent performance once coordination, governance, and cost constraints are included.
\vspace{-2mm}
\subsection{Quantum-Resilient Trust, Identity, and Governance}

The most immediate intersection between quantum computing and agentic AI lies in cryptographic trust. Agentic systems are persistent by design. They store memory, credentials, and decision histories over long time horizons. This makes them especially vulnerable to harvest-now decrypt-later attacks once quantum adversaries become practical~\cite{Clark2025QuantumAgenticAIDataSecurity,Khoury2025AIQuantumEmergingRisks}.

Industry and policy analyses emphasize that agentic AI amplifies cryptographic risk because agents act without human supervision~\cite{Hidary2025NonHumanIdentities}. Non-human identities, delegated authority, and autonomous credentials introduce failure modes that do not exist in user-driven systems. Several analyses argue that quantum-resistant cryptography must be embedded early rather than retrofitted later~\cite{Kataria2025AgenticAIQuantumResistant,Mitchell2025QuantumAgenticAI}.
Academic engagement remains limited. The Aegis Protocol is a notable exception~\cite{aegis2025}. It proposes embedding security controls directly into agent workflows. This is a structural insight. Security is treated as part of agency rather than as an external layer. However, the protocol does not fully specify how post-quantum cryptography interacts with agent memory updates, learning processes, or multi-agent coordination.

Foresight studies examining artificial intelligence, quantum computing, and cybersecurity at a societal scale reinforce this concern~\cite{ai_quantum_2040}. They anticipate convergence, but do not provide design-level guidance for autonomous systems. The unresolved issue is not whether quantum resistance is required. The issue is how to preserve agent autonomy, persistence, and coordination while cryptographic assumptions evolve. Most agentic AI frameworks assume stable cryptography. Most post-quantum cryptography research assumes short-lived or stateless clients. Autonomous agents violate both assumptions. Current literature does not yet resolve this mismatch.

\vspace{-3mm}
\section{Prototype Agentic AI Implementations for Cybersecurity}
\label{sec:implementations}

In addition to surveying frameworks and literature, we prototyped several minimal implementations to illustrate the feasibility of agentic AI in cybersecurity. These are designed for safety and reproducibility, relying on simulated configurations, synthetic logs, and lightweight local models (e.g., Mistral via Ollama). 

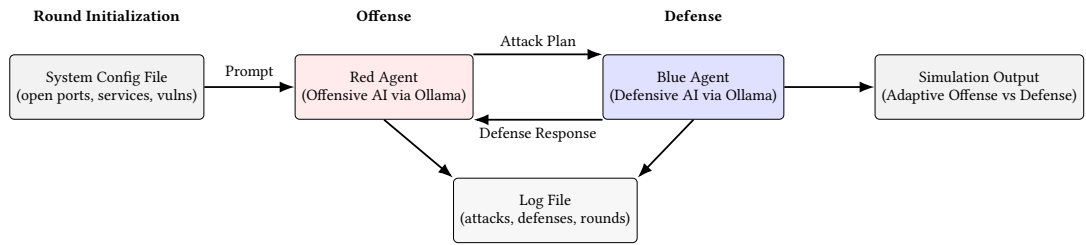
\begin{figure*}[t]
\centering
\resizebox{0.95\linewidth}{!}{%
\begin{tikzpicture}[
    font=\footnotesize, >=Latex,
    node distance=9mm and 14mm,
    box/.style={draw, rounded corners=2pt, align=center, minimum width=26mm, minimum height=10mm},
    process/.style={box, fill=black!3},
    agent/.style={box, fill=blue!5},
    redagent/.style={box, fill=red!8},
    blueagent/.style={box, fill=blue!12},
    io/.style={box, fill=gray!10}
]

\node[io] (input) {System Config File\\ (open ports, services, vulns)};

\node[redagent, right=of input] (red) {Red Agent\\ (Offensive AI via Ollama)};

\node[blueagent, right=2cm of red] (blue) {Blue Agent\\ (Defensive AI via Ollama)};

\node[process, below=of red, xshift=2.5cm] (log) {Log File\\ (attacks, defenses, rounds)};

\draw[->, thick] (input.east) -- (red.west) node[midway, above]{Prompt};
\draw[->, thick] (red.north east) -- (blue.north west) node[midway, above]{Attack Plan};

\draw[->, thick] (blue.south west) -- (red.south east) node[midway, below]{Defense Response};


\draw[->, thick] (red.south) -- (log.north west);
\draw[->, thick] (blue.south) -- (log.north east);

\node[io, right=of blue] (out) {Simulation Output\\ (Adaptive Offense vs Defense)};
\draw[->, thick] (blue.east) -- (out.west);

\node[above=4mm of input] {\textbf{Round Initialization}};
\node[above=4mm of red] {\textbf{Offense}};
\node[above=4mm of blue] {\textbf{Defense}};

\end{tikzpicture}
}
\caption{Workflow of adaptive red–blue simulation. System configuration is ingested, red agent generates attack strategies, blue agent responds with defenses, and exchanges are logged. Loop continues for multiple rounds using Ollama with a lightweight LLM (Mistral).}
\label{fig:implementation_pipeline}
\end{figure*}
\vspace{-3mm}
\subsection{Adaptive Offensive and Defensive Agents}
\label{subsec:adaptive_agents}

Adaptive offensive and defensive agents simulate adversarial cyber behaviors through a red-versus-blue paradigm. A red team agent emulates offensive tactics such as reconnaissance, exploitation, and privilege escalation, while a blue team agent dynamically counters with defense-in-depth strategies. 

We implemented a local interaction loop using a lightweight LLM (Mistral) via the Ollama framework. Red agent was tasked with analyzing a structured system configuration file having data such as open ports, service versions, and known vulnerabilities. Based on this input, red agent generated an attack plan exploiting an Apache RCE vulnerability. Blue agent then reviewed the proposed attack and responded with a multi-layered defense strategy. In the second round, the red agent adapted with evasive and multi-vector tactics such as polymorphic malware, social engineering, and lateral movement, while the blue agent escalated with supply chain monitoring, advanced threat detection, and user behavior analytics. 
All offensive and defensive actions in each round were autonomously generated by the agents. Human user did not specify attack steps, defensive rules, or mitigation tactics. The user provided only two inputs: (i) a static system configuration file describing the environment (e.g., open ports, service versions, known vulnerabilities), and (ii) a high-level role instruction defining 
the agent’s objective (offensive for red, and defensive for blue agent). Red agent independently selected vulnerabilities, attack vectors, and post-exploitation strategies based on the system state. The blue agent was not pre-programmed with fixed defenses or explicit security policies; it reacted solely to the red agent’s proposed actions and inferred appropriate countermeasures with 
a general goal of preventing compromise and data exfiltration. In Round~2, both agents adapted their strategies autonomously in response to the opponent’s prior behavior, without additional user guidance. Implementation workflow is represented in Figure~\ref{fig:implementation_pipeline}, and explained below.

\begin{quote}
\textbf{Round 1 -- Red Agent:}  
``Use the known Apache 2.4.49 RCE vulnerability to gain unauthorized access. Exploit it using Metasploit, establish a reverse shell, exfiltrate data, and delete logs to remain undetected.''

\textbf{Round 1 -- Blue Agent:}  
``Patch the Apache server, restrict access to port 80 via the firewall, monitor logs for shell activity, deploy IDS/IPS to detect RCE patterns, and trigger incident response on anomalies.''

\medskip

\textbf{Round 2 -- Red Agent:}  
``Use a zero-day exploit to bypass patching, initiate a phishing campaign to capture credentials, deploy polymorphic malware to evade detection, move laterally using pass-the-hash, and launch a DDoS to divert attention from exfiltration.''

\textbf{Round 2 -- Blue Agent:}  
``Deploy advanced threat protection (ATP) to detect behavior-based anomalies, train employees to resist phishing, implement network segmentation to contain lateral movement, and secure third-party dependencies to defend against supply chain compromise.''
\end{quote}

This simulation illustrates adaptive reasoning capabilities of agentic AI beyond static prompts. Though textual, the loop mimics realistic escalation, defense posture tuning, and adversarial persistence. This could be extended  with log parsing, external tool access (e.g., Nmap), and deployment into interactive cyber ranges for adversarial resilience testing.

\vspace{-3mm}
\subsection{SOC Triage Agent}
\label{subsec:soc_triage}

\begin{figure*}[t]
\centering
\resizebox{0.95\linewidth}{!}{%
\begin{tikzpicture}[
    font=\footnotesize, >=Latex,
    node distance=10mm and 14mm,
    box/.style={draw, rounded corners=2pt, align=center, minimum width=30mm, minimum height=10mm},
    io/.style={box, fill=gray!10},
    process/.style={box, fill=blue!5},
    baseline/.style={box, fill=green!10},
    agent/.style={box, fill=red!10},
    gov/.style={box, fill=orange!10},
    result/.style={box, fill=black!5}
]

\node[io] (alerts) {Synthetic Alerts\\ (Netflow/IDS events)};
\node[io, below=of alerts] (enrich) {Enrichment Data\\ (Assets, Threat Intel)};

\node[baseline, right=25mm of alerts] (base) {Baseline Rules\\ (heuristics)};
\node[agent, below=of base] (llm) {LLM Agent\\ (Mistral via Ollama)};

\node[gov, right=25mm of base, yshift=-10mm] (policy) {Policy Gate\\ (read-only enforcement)};

\node[result, right=25mm of policy, yshift=10mm] (audit) {Audit Log};
\node[result, right=25mm of policy, yshift=-10mm] (results) {Decisions + Metrics\\ (Precision, Recall, MTTR)};

\draw[->, thick] (alerts.east) -- (base.west);
\draw[->, thick] (enrich.east) -- (base.west);
\draw[->, thick] (alerts.east) -- (llm.west);
\draw[->, thick] (enrich.east) -- (llm.west);

\draw[->, thick] (base.east) -- (policy.west);
\draw[->, thick] (llm.east) -- (policy.west);

\draw[->, thick] (policy.east) -- (audit.west);
\draw[->, thick] (policy.east) -- (results.west);

\end{tikzpicture}
}
\caption{SOC triage agent workflow. Synthetic alerts and enrichment data are processed by baseline rules and an LLM agent. All actions pass through a policy gate, with results logged and evaluated for precision, recall, F1, and MTTR.}
\label{fig:soc_triage_pipeline}
\end{figure*}
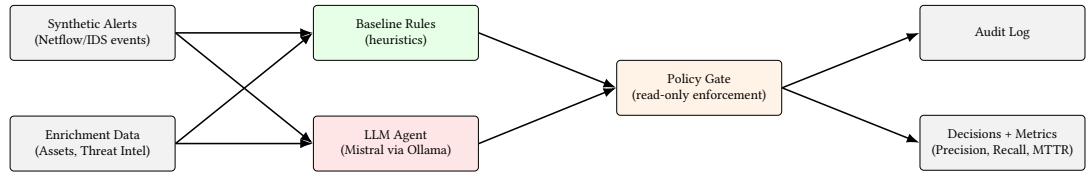

SOC triage agents automate Tier-1 alert handling by filtering, enriching, and prioritizing events. We compare a deterministic rule-based baseline with an LLM-assisted agent to evaluate whether agentic AI can improve recall and reduce mean time to triage (MTTR) in a controlled synthetic setting.
The workflow ingests synthetic alerts (Apache Struts RCE, SSH brute force, SQL injection, malware hash detection, suspicious user agent, and port scan) and enriches them with asset criticality and threat intelligence. The baseline applies static heuristics, while the agentic version queries a local LLM (Mistral via Ollama) to output structured decisions (\texttt{escalate} or \texttt{close}) with rationales. A governance layer enforces read-only actions and logs all decisions. The end-to-end pipeline is shown in Figure~\ref{fig:soc_triage_pipeline}.

Example agent decisions include escalating a high-severity Apache Struts RCE on a critical web asset, closing a suspicious user agent from a CI system with no malicious indicators, and escalating a known malicious hash on a critical HR endpoint.
Evaluation on six alerts with ground-truth labels shows that the baseline achieved perfect precision but lower recall (0.75), while the LLM agent achieved perfect precision, recall, and F1, and reduced simulated MTTR from 7.5 to 6.0 minutes (Table~\ref{tab:soc_triage_results}). All 12 agent actions were policy-compliant and fully logged. This result indicates that LLM-based triage can reduce missed detections and improve efficiency without sacrificing precision.








\vspace{-3mm}
\subsection{Anomaly and IDS Agent}
\label{subsec:anomaly_ids}

Anomaly detection and intrusion detection systems (IDS) provide a natural testbed for agentic AI because they require both statistical sensitivity and contextual reasoning. Traditional approaches rely on rules or unsupervised detectors (e.g., z-score thresholds or Isolation Forests) to flag traffic spikes. However, these methods often generate false positives by failing to distinguish benign periodicity from malicious anomalies. Agentic AI can augment this process by combining lightweight retrieval with natural language reasoning, thereby improving interpretability and triage efficiency.

\begin{figure*}[t]
\centering
\resizebox{0.95\linewidth}{!}{%
\begin{tikzpicture}[
    font=\footnotesize, >=Latex,
    node distance=10mm and 16mm,
    box/.style={draw, rounded corners=2pt, align=center, minimum width=28mm, minimum height=10mm},
    io/.style={box, fill=gray!10},
    baseline/.style={box, fill=green!10},
    agent/.style={box, fill=blue!10},
    gov/.style={box, fill=orange!10},
    result/.style={box, fill=black!5}
]

\node[io] (input) {Synthetic NetFlow Windows\\ (bytes/sec, pkts/sec, SYN rate, failed conn)};
\node[io, below=of input] (rag) {RAG Corpus\\ (benign vs malicious patterns)};

\node[baseline, right=20mm of input] (zscore) {Baseline Detector\\ (z-score, IForest)};

\node[agent, below=of zscore] (llm) {LLM Agent (Mistral)\\ + RAG notes + rationale};

\node[gov, right=20mm of zscore, yshift=-10mm] (policy) {Policy Gate\\ (read-only enforcement)};

\node[result, right=20mm of policy, yshift=10mm] (log) {Audit Log\\ (prompts, responses)};
\node[result, right=20mm of policy, yshift=-10mm] (out) {Decisions + Metrics\\ (AUROC, F1, rationale scores)};

\draw[->, thick] (input.east) -- (zscore.west);
\draw[->, thick] (input.east) -- (llm.west);
\draw[->, thick] (rag.east) -- (llm.west);

\draw[->, thick] (zscore.east) -- (policy.west);
\draw[->, thick] (llm.east) -- (policy.west);

\draw[->, thick] (policy.east) -- (log.west);
\draw[->, thick] (policy.east) -- (out.west);

\end{tikzpicture}
}
\caption{Anomaly/IDS agent workflow. Synthetic NetFlow windows are analyzed by both a baseline detector (z-score, IForest) and an LLM agent with retrieval-augmented notes. All actions pass through a policy gate and are recorded in an audit log. Outputs include anomaly decisions, metrics (AUROC, F1), and explanation usefulness scores.}
\label{fig:anomaly_ids_pipeline}
\end{figure*}
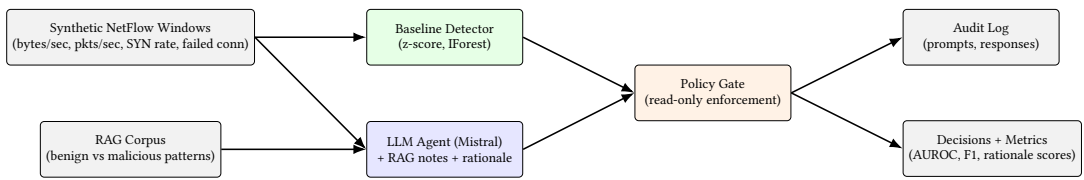

We implemented a micro-demo using synthetic NetFlow-like time windows. Each window contained summary features such as bytes-per-second, packets-per-second, SYN rate, and failed connection rate. The baseline detector applied a z-score rule and Isolation Forest to classify anomalous windows. In contrast, the agentic IDS queried a local LLM (Mistral via Ollama) with both the raw statistics and a small retrieval-augmented note set describing benign patterns (e.g., periodic backups) and malicious signatures (e.g., volumetric spikes). The agent produced JSON-structured outputs with binary labels and concise rationales. The overall workflow is shown in Figure~\ref{fig:anomaly_ids_pipeline}, as explained below:


\begin{quote}
\textbf{Window 0 (LLM Agent):}  
``The window exhibits volumetric anomalies due to sharp spikes in bytes\_per\_sec and pkts\_per\_sec, which contradicts the note of benign periodicity. Therefore, label = 1 (anomalous).''

\medskip

\textbf{Window 19 (LLM Agent):}  
``While the traffic rates are elevated, the presence of benign periodicity suggests normal scheduled activity such as backups. Therefore, label = 0 (normal).''
\end{quote}

The evaluation compared baseline detectors and the agentic model on synthetic windows with ground-truth labels. As shown in Table~\ref{tab:ids_results}, the baseline achieved reasonable AUROC but produced false positives on periodic traffic. The agent maintained high recall while improving F1 and providing human-readable rationales. A small rubric was also applied to evaluate explanation usefulness (consistency, specificity), where the agent achieved an average score of 0.83.


This experiment shows how agentic IDS prototypes can blend statistical detection with explainable reasoning. Unlike traditional detectors that output only binary alerts, the agent provided concise justifications grounded in retrieved knowledge, making outputs more useful for analysts. Future work could integrate richer traffic features (e.g., flow durations, entropy measures) and evaluate robustness in adversarial settings or interactive SOC ranges.







\begin{table*}[!t]
\centering
\small

\begin{minipage}[t]{0.48\textwidth}
\centering
\renewcommand{\arraystretch}{1.2}
\setlength{\tabcolsep}{5pt}
\caption{Performance of SOC triage baseline vs.\ LLM-assisted agent on synthetic alerts.}
\label{tab:soc_triage_results}
\rowcolors{2}{gray!10}{white}
\resizebox{\textwidth}{!}{%
\begin{tabular}{p{0.26\linewidth} p{0.16\linewidth} p{0.16\linewidth} p{0.08\linewidth} p{0.30\linewidth}}
\toprule
\rowcolor{gray!35}
\textbf{System} & \textbf{Precision} & \textbf{Recall} & \textbf{F1} & \textbf{Sim.\ MTTR (min)} \\
\midrule
Baseline (rules) & 1.00 & 0.75 & 0.86 & 7.5 \\
Agent (Mistral)  & 1.00 & 1.00 & 1.00 & 6.0 \\
\bottomrule
\end{tabular}}
\end{minipage}
\hfill
\begin{minipage}[t]{0.48\textwidth}
\centering
\renewcommand{\arraystretch}{1.2}
\setlength{\tabcolsep}{5pt}
\caption{Performance of baseline vs.\ LLM-assisted anomaly/IDS agent on synthetic NetFlow windows.}
\label{tab:ids_results}
\rowcolors{2}{gray!10}{white}
\resizebox{0.9\textwidth}{!}{%
\begin{tabular}{p{0.45\linewidth} p{0.25\linewidth} p{0.25\linewidth}}
\toprule
\rowcolor{gray!35}
\textbf{System} & \textbf{AUROC} & \textbf{F1} \\
\midrule
Baseline (z-score + IForest) & 0.87 & 0.74 \\
Agent (Mistral)              & 0.93 & 0.81 \\
\bottomrule
\end{tabular}}
\end{minipage}

\end{table*}

\vspace{-3mm}
\section{Directions of Future Research}
\label{future_research_directions}
Agentic AI security remains an open research. Existing work identified many threats, but defensive solutions are still novice \cite{owasp2025agentic,securitythreats2025}. Future research should focus on closing the gap between agent capabilities and security guarantees.

\noindent
\textbf{Input inspection and control:} 
Research should improve automatic inspection of user inputs. Current defenses against prompt injection and jailbreak attacks are incomplete and inefficient \cite{securitythreats2025,owasp2025agentic}. Many systems rely on heuristic filters or offline analysis which do not scale well to real time agent execution. Research should develop lightweight and adaptive inspection mechanisms that operate during agent interaction. These mechanisms should distinguish between instructions and data with high reliability. They should also adapt to multistep and multimodal inputs \cite{Evtimov2025WASP,Wang2025AgentVigil}.

\noindent
\textbf{Transparency of internal execution:} 
Internal execution of agents is difficult to observe. Reasoning chains, planning steps, and tool calls are often hidden from auditors, limiting timely detection of unsafe behavior \cite{he2025sentinel}. Research should design methods to expose internal states without leaking sensitive information. Structured logging and execution traces are promising directions which can support runtime verification and post hoc auditing of agent behavior \cite{gaas2025,horus2025}.

\vspace{-1mm}
\noindent
\textbf{Robust planning and reasoning:} 
Planning errors can amplify small mistakes into severe failures. Current planning structures lack formal guarantees. Future work should study error propagation in multistep reasoning. Research should also explore constrained planning methods that enforce safety rules throughout execution. Combining language models with formal constraints or verifiable policies remains an open challenge \cite{horus2025,owasp2025agentic}.

\noindent
\textbf{Secure interaction with environments:} 
Agents increasingly interact with dynamic and untrusted environments. Indirect prompt injection and manipulated feedback remain serious risks \cite{securitythreats2025,owasp2025agentic}. Future research should develop stronger isolation between agent instructions and external data. Secure sandboxing and environment validation should be studied across deployment settings. This includes cloud systems, simulated environments, and physical systems.

\noindent
\textbf{Multi agent coordination security:} 
Multi agent systems introduce new attack surfaces as agents may collude or propagate errors. Competitive agents may deceive or manipulate each other. Existing defenses are limited and often reduce system efficiency \cite{he2025sentinel}. Future work should study secure coordination protocols for agent communication. Research should also examine how trust, verification, and accountability can be enforced across agents \cite{blocka2a2025,nanda2025,zerotrust2025,loka2025}.

\noindent
\textbf{Memory integrity and privacy:} 
Agent memory is a critical vulnerability. Short term memory limits reasoning. Long term memory can be poisoned or leaked \cite{owasp2025agentic,securitythreats2025}. Current defenses focus on model level protections rather than memory systems. Future research should design secure memory architectures for agents. These architectures should support validation, versioning, and access control. Privacy preserving retrieval methods are also needed \cite{li2025safeflow,he2025sentinel}.

\noindent
\textbf{Unified evaluation and benchmarks:}
Security evaluation of agents lacks standardized benchmarks. Existing datasets focus on isolated attacks or single agents. Future research should develop comprehensive benchmarks that cover perception, reasoning, action, interaction, and memory. These benchmarks should reflect real world deployment scenarios. They should also support reproducible and comparable evaluation of defenses \cite{Evtimov2025WASP,Wang2025AgentVigil,bountybench2025}.

\noindent
\textbf{Governance and deployment assurance:}
Technical defenses alone are not sufficient. Agentic systems operate under policy, legal, and organizational constraints. Future research should integrate governance mechanisms into agent design \cite{gaas2025,aegis2025}. This includes policy enforcement, auditing, and human oversight. Assurance frameworks that combine technical and organizational controls remain an open research direction \cite{loka2025,nanda2025,blocka2a2025,zerotrust2025}.

\vspace{-2mm}
\section{Conclusion}
\label{sec:conclusion}

Agentic AI marks a shift from static inference to goal-directed systems that reason, act, and adapt over time. By integrating memory, tool use, and autonomy, these systems enable new cybersecurity capabilities across monitoring, response, intelligence, and training. At the same time, they introduce novel risks that arise from persistent state, execution authority, and multi-agent coordination.
Across the literature, a consistent tradeoff emerges. Greater autonomy improves speed and adaptability but reduces predictability, auditability, and control. Security failures stem from system-level interactions among perception, reasoning, action, memory, and identity rather than from model inference alone. Existing frameworks, benchmarks, and governance approaches address parts of this problem but remain fragmented.

As agentic AI becomes embedded in critical cybersecurity infrastructure, security and governance must be treated as foundational design requirements. Progress will depend on bounded autonomy, enforceable execution controls, continuous assurance, and human oversight. Addressing these challenges is essential to realizing the benefits of agentic AI while limiting misuse and systemic risk.

\bibliographystyle{ACM-Reference-Format}
\bibliography{references}

\end{document}